\documentclass[preprint,12pt]{elsarticle}

\usepackage{float}

\usepackage{amssymb}
\usepackage{amsmath}

\usepackage[linesnumbered,ruled,vlined]{algorithm2e}
\usepackage{mathtools}
\usepackage{subcaption}

\usepackage{bm}

\usepackage{lipsum}
\usepackage{fancyhdr}

\journal{Physical Communication}

\begin{document}

\begin{frontmatter}



\title{
	Resilient Signal Reflection under CSI Perturbations: A Robust Approach for Secure RIS Communication}
 \author[label1]{Mahdi Shamsi}
 \author[label2]{ Hadi Zayyani}
 \author[label1]{Farokh Marvasti}
 \affiliation[label1]{organization={Multimedia and Signal processing Lab (MSL),
 				EE Department of Sharif University of Technology},
             city={Tehran},
             country={Iran}}

 \affiliation[label2]{organization={Department of Electrical and Computer Engineering,
 				Qom University of Technology},
             city={Qom},
             country={Iran}}


\begin{abstract}
Reconfigurable Intelligent Surfaces (RIS) have emerged as a transformative technology in wireless communication, enabling dynamic control over signal propagation. This paper tackles the challenge of mitigating Channel State Information (CSI) perturbations in RIS-aided systems, particularly for secure communication scenarios. Leveraging a first-order approximation technique, we develop a robust approach that strengthens the resilience of RIS configurations against CSI imperfections. The study considers both untrusted user interception and stealth radar applications, focusing on optimizing signal reflection and transmission in the presence of eavesdroppers. Simulation results demonstrate notable gains in security and efficiency while maintaining low computational complexity. By extending the stability range, the proposed method updates RIS elements using only a few matrix-vector multiplications, eliminating the need for repeated inverse or pseudo-inverse computations under small channel perturbations. Additionally, the framework provides a baseline for quantifying algorithmic sensitivity to CSI variations. Overall, the findings underscore the potential of RIS to enable secure and reliable communication in next-generation networks such as 6G.
\end{abstract}

%

\begin{keyword}
		Reconfigurable Intelligent Surface, Secure Communication, Imperfect CSI, Active RIS.
\end{keyword}

\end{frontmatter}


	\section{Introduction}
\label{sec:introduction}
Reconfigurable Intelligent Surface (RIS)\footnote{Also referred to as Intelligent Reflecting Surfaces, the terms are used interchangeably.} systems and technologies are innovative structures that enhance wireless communication by controlling electromagnetic waves \cite{siddiqi2022reconfigurable, jian2022reconfigurable}. Typical RIS comprises passive and active elements that can adjust parameters such as phase and amplitude to improve signal propagation \cite{wu2019intelligent}. This technology enhances communication performance by directing signals to specific users or mitigating interference \cite{yildirim2022channel}, with applications across smart cities \cite{kisseleff2020reconfigurable}, wireless networks \cite{10460579}, IoT \cite{das2023comprehensive}, and 6G \cite{shi2024ris}.

Recent progress in Reconfigurable Intelligent Surfaces (RIS) has greatly advanced wireless communication by enabling dynamic control of the propagation environment. RIS-assisted ISAC systems improve radar detection and communication reliability, while active and STAR-RIS architectures address limitations of passive RIS by enhancing coverage, security, and secrecy performance through advanced optimization methods \cite{jin2025beamforming, WANG2025102829, zhong2025active}. Hybrid RIS designs combining active and passive elements further boost spectral efficiency and reduce error rates in full-duplex systems \cite{shafrin2025hybrid}, and robust resource allocation strategies ensure secure operation even under imperfect channel knowledge \cite{yue2024robust}. Collectively, these developments underscore the potential of RIS technologies; passive, active, STAR, and hybrid; as key enablers of next-generation secure and high-performance wireless networks.



Low Probability of Intercept (LPI) and Low Probability of Detection (LPD) are critical in secure radar and communication systems
\cite{du2025joint,carroll2017chaos, cui19, wang20, Liu24, amiriara2023efficient}.
 Techniques for achieving LPI and LPD include RIS-aided communication \cite{cui19, souzani2023physical}, waveform design \cite{li2013new, liu2023lpi, shi2018low}, structured radio beams \cite{zhou2019low}, and antenna arrays \cite{wang2021lpi, liu2023lpic, zhao2024low}. RIS is particularly promising for secure next-generation communication and radar systems \cite{cui19, shen19, wang20, Liu24}. It can effectively mask signals from untrusted users (Fig. \ref{fig:config2} and \ref{fig:config3}) and protect stealth radars from Electronic Support Machines (ESM) (Fig.~\ref{fig:config4}).

Enhancing system security is a critical aspect of modern wireless communication systems. In \cite{sahu2025secure}, a blockchain-based Internet of Vehicles (IoV) architecture is proposed, incorporating Physically Unclonable Functions (PUFs) and bloom filters to enable lightweight, privacy-preserving, and verifiable authentication for Electric Vehicle (EV) networks. To further strengthen vehicular network security, \cite{kishore2025secure} presents a Digital Twin-based Lightweight Hashing Blockchain Cryptography (DT-LHBC) scheme for VANETs, which demonstrates high resilience against various attacks while maintaining computational efficiency and scalability in dynamic environments. The integration of RIS into secure communication systems was initially explored in \cite{cui19}, where joint optimization of transmit beamforming and RIS phase shifts was employed to maximize the secrecy rate. Subsequent works have extended this framework to secure MISO channels \cite{wang20}, proposed efficient transmission and phase shift optimization algorithms \cite{shen19, Feng21}, and applied RIS to enhance physical layer security in Non-Orthogonal Multiple Access (NOMA) systems \cite{souzani2023physical} and joint radar-communication platforms \cite{Liu24, zhang2023deep}. These advancements collectively demonstrate the evolving role of RIS and blockchain in fortifying secure wireless communications.

In this paper, unlike previous works that assume unperturbed knowledge of the Channel State Information (CSI) for either the untrusted user (Eve) or legitimate user (Bob) \cite{cui19, shen19}, or omit CSI entirely \cite{wang20, zhang2023deep}, we address CSI uncertainty in channel scenarios. We employ first-order perturbation analysis to tackle practical issues and present a unified model for both communication and radar scenarios. The analysis primarily employs the Least Squares Solution (LSS) for sensitivity analysis, and we provide a thorough investigation along with regularization techniques. Despite its simplicity, the least squares method remains a fundamental tool for performance optimization, as evidenced by its continued relevance in contemporary research areas such as subspace learning \cite{chen2022jointly} and signal estimation in transmission lines \cite{aljweer2024new}.

Simulation results indicate that the LSS approach (pseudo-inverse method - "pinv") can effectively reduce the signal level to match the noise level when perfect CSI is available. For instance, it has been demonstrated that this method can achieve a 40dB reduction in signal level at a Signal-to-Noise Ratio (SNR) of 40dB. However, it is important to note that such performance is highly sensitive to channel perturbations, which can be categorized from two perspectives:
\begin{enumerate}
	\item imperfect CSI due to non-ideal channel estimation,
	\item variations in CSI resulting from the movements of the transmitter, receiver, or even the RIS itself (for example, when mounted on a UAV).
\end{enumerate}
The first aspect leads to deviations from the optimal solution, while the second necessitates continuous adjustment to maintain performance.

To enhance stability across both scenarios, we employ Least Absolute Shrinkage and Selection Operator (LASSO) and Ridge regularization techniques \cite{hastie2009elements}. We derive a first-order approximation to capture changes in the RIS element values concerning channel perturbations. This approach not only addresses the drift in parameters but also illustrates how regularization can enhance performance by reducing sensitivity and expanding the stability range. Furthermore, utilizing this method enables edge computing devices to avoid excessive calculations of optimal values in cases of relatively minor changes, allowing them to rely on the computationally efficient first-order approximation.

Additionally, we tackle the passiveness constraint by employing Projected Gradient Descent (PGD) and solve the complex LASSO problem using the Iterative Shrinkage Thresholding Algorithm (ISTA) \cite{beck2009fast}.

The novelty and contributions of this paper include: 
\begin{itemize}
	\item Formulating and addressing the problem of imperfect CSI,
	\item Deriving a first-order approximation to quantify result deviations,
	\item Implementing regularizers (Ridge and LASSO) to enhance stability and robustness against varying CSI due to both non-ideal channel estimation and dynamic changes,
	\item Simulating the sensitivity of the performance of the algorithms with respect to channel perturbations and comparing it with the first-order approximation.
\end{itemize}

The paper is organized as follows:
\newline The literature survey is covered in the introduction. Section~\ref{sec: sysmodel} presents the system and signals model, followed by problem formulation. The LSS using the pseudo-inverse is discussed in Section~\ref{sec: LS}, while Section~\ref{sec: reg} explores the regularized LSS. Simulation results are provided in Section~\ref{sec: sim}, and conclusions are drawn in Section~\ref{sec: con}.

\section{The System and Signals model}
\label{sec: sysmodel}
\begin{figure}[bth!]
	\centering
	\includegraphics[trim={0 0 0 0}, scale=0.25]{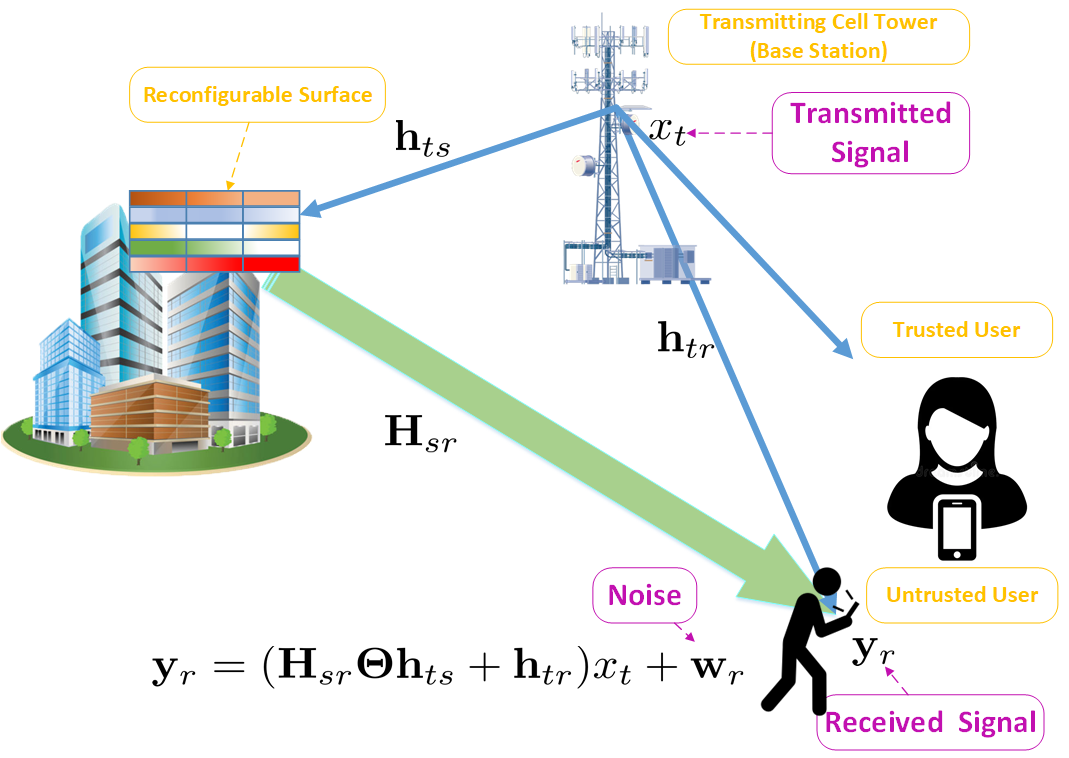}
	\caption{Masking an Untrusted User from the Base Station.}
	\label{fig:config2}
\end{figure}

\begin{figure}[bth!]
	\centering
	\includegraphics[trim={0 0 0 0}, scale=0.25]{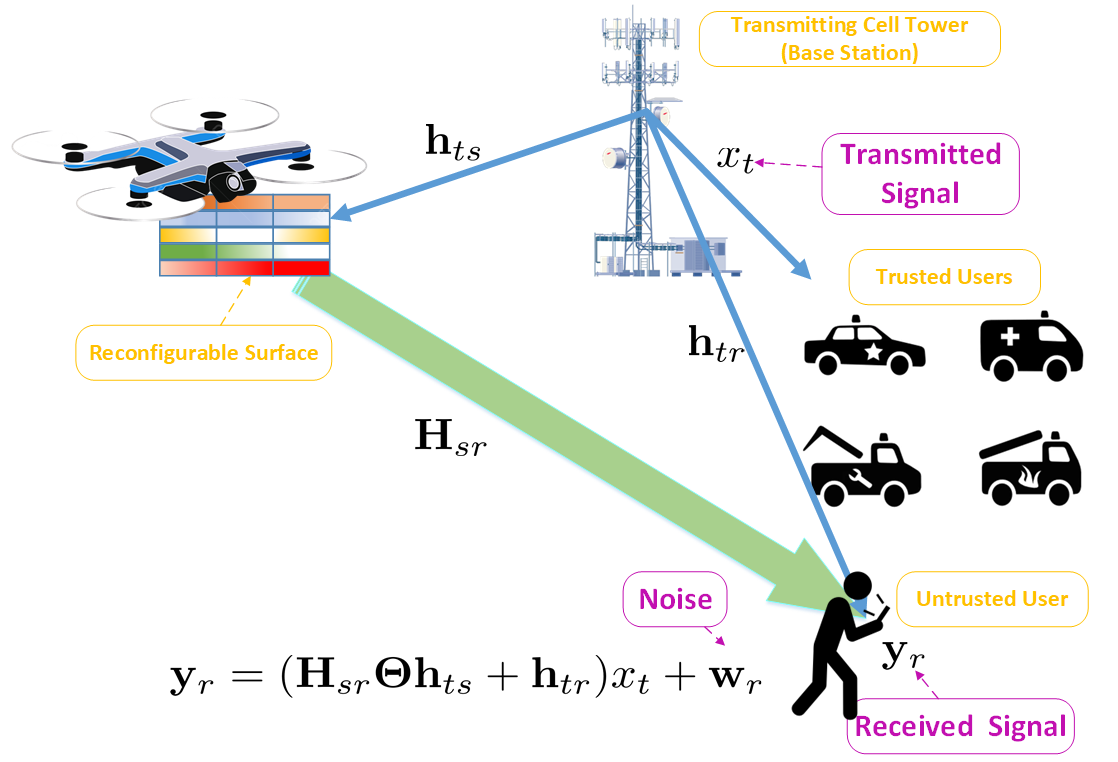}
	\caption{Masking an Untrusted User, RIS mounted on a UAV.}
	\label{fig:config3}
\end{figure}

\begin{figure}[bth!]
	\centering
	\includegraphics[trim={0 0 0 0}, scale=0.275]{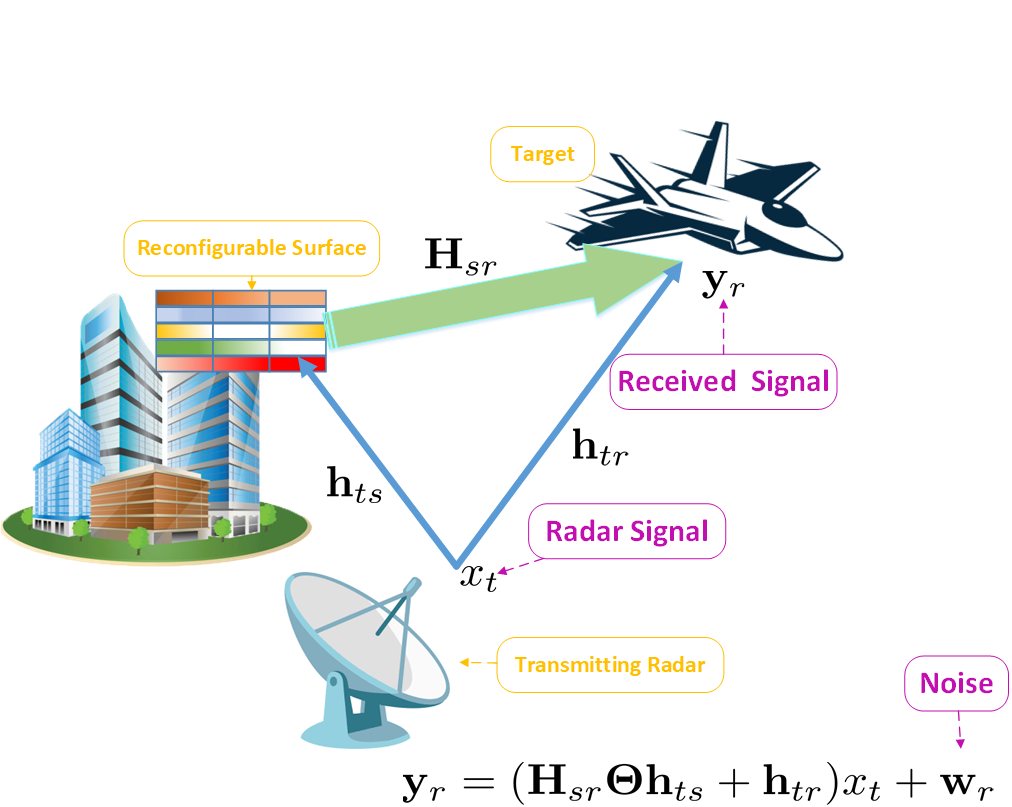}
	\caption{Stealth Radar.}
	\label{fig:config4}
\end{figure}

In this section, we unify three scenarios depicted in Fig.~\ref{fig:config2}, \ref{fig:config3}, and \ref{fig:config4}. These typical scenarios cover possible situations where channel perturbations can arise due to the movement of participating components (including nomadic RIS scenarios). The first scenario represents a standard RIS-aided configuration; the second scenario aligns with a configuration commonly adopted in Disaster Management applications \cite{10582499}; and the third scenario illustrates a variant designed for electromagnetic stealth purposes \cite{10575930}.

The first two scenarios involve communication setups where an untrusted user (Eve) attempts to intercept the signal intended for an authorized end user. In contrast, the third scenario is a radar application aimed at maintaining stealth from the perspective of an ESM target.

We adopt the signal model from \cite{siddiqi2022reconfigurable}:
\begin{equation}
	\mathbf{y}_r=(\mathbf{H}_{sr}\mathbf{\Theta}\mathbf{h}_{ts}+\mathbf{h}_{tr})x_t+\mathbf{w}_r,
\end{equation}
where $\mathbf{h}_{tr}$ represents the channel between the transmitter and the untrusted user, $\mathbf{H}_{sr}$ is the channel between the RIS and the untrusted receiver, $\mathbf{h}_{ts}$ denotes the channel between the transmitter and the RIS, and $\mathbf{\Theta}$ is the RIS phase and amplitude adjustment matrix.

We explore both passive and active RIS configurations, as active RIS technologies can overcome the limitations of multiplicative fading effect \cite{zhang2022active} and enhance secure transmission by jointly optimizing phase and amplitude adjustments \cite{dong2021active}. For more on channel and path loss modeling, see \cite{tang2020wireless, tang2022path, yildirim2022channel}.

For simplicity, the key variables are redefined as follows:
\begin{equation}
	\mathbf{b} = \mathbf{h}_{tr},\;\mathbf{A} = \mathbf{H}_{sr},\;\mathbf{c}  = \mathbf{h}_{ts}, \; \mathbf{D} = \mathbf{\Theta}.
\end{equation}
Thus, given a receiver with $m$ antennas and an RIS with $n$ elements, we have the following properties:
\begin{equation}
	\mathbf{b} \in \mathbb{C}^m,\;\mathbf{A} \in \mathbb{C}^{m \times n},\;\mathbf{D} \in \mathbb{C}^{n \times n},\;\mathbf{c} \in \mathbb{C}^n.
	\label{eq:params}
\end{equation}
Substituting these into the original model yields:
\begin{equation}
	\mathbf{y}_r=  (\mathbf{b} + \mathbf{A}\mathbf{D}\mathbf{c})\mathbf{x}_t+\mathbf{w}_r,
	\label{eq:main model}
\end{equation}
where $\mathbf{D} = \text{diag}(d_1, d_2, \dots, d_n)$ is an $n \times n$ diagonal complex matrix. The vectors $\mathbf{y}_r$, $\mathbf{x}_t$, and $\mathbf{w}_r$ represent the target's received signal, the transmitter's signal, and the receiver's additive noise, respectively.

We want to minimize the received signal level:
\begin{equation}
	\|\mathbf{y}_r\|^2= \| (\mathbf{b} + \mathbf{A}\mathbf{D}\mathbf{c})\mathbf{x}_t \|^2 +\|\mathbf{w}_r\|^2,
\end{equation}
w.r.t the diagonal elements of $\mathbf{D}$.
In other words, due to the independence of the signal and noise, we want to minimize
\begin{equation}
	f(\mathbf{d},\lambda) = \| \mathbf{b} + \mathbf{A}_c\mathbf{d} \|^2+R(\mathbf{d},\lambda),
	\label{eq:optimization}
\end{equation}
w.r.t $\mathbf{d} =\text{vec}(d_1, d_2, \dots, d_n)$, where $R(\mathbf{d},\lambda)$ is a regularization term with parameter $\lambda$.
It is also desired to achieve
\begin{equation}
	\mathbf{d}_{\text{opt}} \triangleq \arg\min_\mathbf{d} f(\mathbf{d},\lambda),
\end{equation}
with the lowest possible signal level $S$ ($S\triangleq\| \mathbf{b} + \mathbf{A}_c\mathbf{d} \|^2$) at a specified sensitivity to the CSI perturbation (denoted by $ \Delta \mathbf{d}_{\text{opt}}\triangleq \mathbf{d}' -\mathbf{d}$, where the $\mathbf{d}'$ is the solution under the perturbed channels).

To analyze the sensitivity of $ \mathbf{d}_{\text{opt}} $ to perturbations in $ \mathbf{A} $, $ \mathbf{b} $, and $ \mathbf{c} $, we will explore how small changes in these variables affect $ \mathbf{d}_{\text{opt}} $. Sensitivity can be approximated by the 1st-order Taylor's expansion with respect to these variables (as $ \delta \mathbf{d}_{\text{opt}}$ which is a first-order approximation of $ \Delta \mathbf{d}_{\text{opt}}$: $\mathbf{d}'_{\text{opt}}  \approx\mathbf{d}_{\text{opt}} +\delta \mathbf{d}_{\text{opt}}$).
Consider small perturbations denoted by:
\begin{itemize}
	\item $ \delta \mathbf{A} $ for the perturbation in $ \mathbf{A} $,
	\item $ \delta \mathbf{b} $ for the perturbation in $ \mathbf{b} $,
	\item $ \delta \mathbf{c} $ for the perturbation in $ \mathbf{c} $,
	\item $ \delta \mathbf{d}_{\text{opt}} $ for the approximated resulting perturbation in $ \mathbf{d}_{\text{opt}} $.
\end{itemize}
Appendices are dedicated to derivation the optimal RIS parameters based on the objective function and accordingly calculating their first-order sensitivity to channels perturbation.

\section{Least-Squares Solution and Pseudo-inverse}
\label{sec: LS}
\subsection{Least-Squares Solution (LSS)}
Considering $R(\mathbf{d},\lambda) = 0$, the system of equations in \eqref{eq:optimization} can be rewritten as a Least-Squares Solution (LSS):
\begin{equation}
	f_{_{LSS}}(\mathbf{d}) = \| \mathbf{b} + \mathbf{A}_c\mathbf{d} \|^2,
\end{equation}
where $ \mathbf{A}_c = \mathbf{A} \odot \mathbf{c} $ and $ \mathbf{d} = \text{vec}(d_1, d_2, \dots, d_n) $ (in other word: $\mathbf{D} = \text{diag}(\mathbf{d})$) with the dimensions expressed in \eqref{eq:params}.
As shown in Appendix \ref{app:LSS}, the solution for $\mathbf{D}$ is\footnote{$\mathbf{A} \odot \mathbf{c}$ denotes element-wise multiplication of each column of $ \mathbf{A} $ by the corresponding element of $ \mathbf{c} $.}\footnote{$ \mathbf{A}^\dagger$ is the complex transpose of $ \mathbf{A}$.}:
\begin{equation}
	\mathbf{d}_{\text{opt, LSS}} = - (\mathbf{A}_c^\dagger \mathbf{A}_c)^{-1} \mathbf{A}_c^\dagger \mathbf{b}.
\end{equation}

As derived in \eqref{eq:sensitivity_reg}, the first-order sensitivity of $ \mathbf{d}_{\text{opt}} $ to perturbations in $ \mathbf{A} $, $ \mathbf{b} $, and $ \mathbf{c} $ is captured by:
\subsubsection{Perturbations in $ \mathbf{A} $}
\begin{align}
	\delta_A \mathbf{d}_{\text{opt, LSS}} = &- (\mathbf{A}_c^\dagger \mathbf{A}_c)^{-1} \left[ (\delta \mathbf{A} \odot  \mathbf{c})^\dagger \mathbf{b}\right]\notag\\
	&- (\mathbf{A}_c^\dagger \mathbf{A}_c)^{-1} (\mathbf{A}_c^\dagger \delta \mathbf{A} \odot  \mathbf{c} + (\delta \mathbf{A} \odot  \mathbf{c})^\dagger \mathbf{A}_c)  \mathbf{d}_{\text{opt, LSS}}.\notag
\end{align}
\subsubsection{Perturbations in $ \mathbf{b} $}
\begin{equation}
	\delta_b \mathbf{d}_{\text{opt, LSS}} = - (\mathbf{A}_c^\dagger \mathbf{A}_c)^{-1} \mathbf{A}_c^\dagger \delta \mathbf{b}.
\end{equation}
\subsubsection{Perturbations in $ \mathbf{c} $}
\begin{align}
	\delta_c \mathbf{d}_{\text{opt, LSS}} = &- (\mathbf{A}_c^\dagger \mathbf{A}_c)^{-1} \left[ (\mathbf{A} \odot \delta \mathbf{c})^\dagger \mathbf{b} \right] \notag\\
	&+ (\mathbf{A}_c^\dagger \mathbf{A}_c)^{-1}\left[ \mathbf{A}_c^\dagger  (\mathbf{A} \odot  \delta\mathbf{c})
	+ ( \mathbf{A} \odot  \delta\mathbf{c})^\dagger \mathbf{A}_c \right]\mathbf{d}_{\text{opt, LSS}}.\notag
\end{align}

The approximated perturbation of $ 	f_{_{LSS}}(\mathbf{d}) $, denoted as $ \delta	f_{_{LSS}}(\mathbf{d})$, will capture the effect of changes in $ \mathbf{A} $, $ \mathbf{b} $, and $ \mathbf{c} $ on the objective function as a notion of increase in the signal level. This gives us the perturbed version of the function $ 	 f_{_{LSS}}(\mathbf{d}) $:
\begin{equation}
	f_{_{LSS}}(\mathbf{d}) + \Delta 	f_{_{LSS}}(\mathbf{d})= \| (\mathbf{b} + \delta \mathbf{b}) + (\mathbf{A}_c + \delta \mathbf{A}_c) \mathbf{d} \|^2.
	\label{eq:fLSS_perturbed}
\end{equation}
As shown in \eqref{eq:f_sens_LSS}, the perturbation is approximated by:
\begin{align}
	\delta 	f_{_{LSS}}(\mathbf{d}) = 2 \Re \left( (\mathbf{b} + \mathbf{A}_c\mathbf{d})^\dagger (\delta \mathbf{b} + (\delta \mathbf{A}\odot\mathbf{c})\mathbf{d}+ (\mathbf{A} \odot\delta\mathbf{c})\mathbf{d}) \right).\notag
\end{align}
\subsection{The Moore-Penrose Pseudo-inverse Approach}
If $ \mathbf{A}_c^\dagger \mathbf{A}_c $ is singular (i.e., non-invertible), then the system $ \mathbf{A}_c^\dagger \mathbf{A}_c \mathbf{d} = -\mathbf{A}_c^\dagger  \mathbf{b} $ does not have a unique solution, or might not have any solution in the strict sense. In such cases, the solution to minimize $ \|\mathbf{A}_c \mathbf{d} + \mathbf{b}\|^2 $ is:
\begin{equation}
	\mathbf{d}_{\text{opt, pinv}} = -\mathbf{A}_c^+ \mathbf{b}.
\end{equation}
Following the approach mentioned in Appendix \ref{app:pinv}, this gives the minimum-norm least-squares solution when the exact solution cannot be obtained due to the singularity of $ \mathbf{A}_c^\dagger \mathbf{A}_c $.

Thus, as derived in \eqref{eq:f_sens_pinv}, the change in the solution $\mathbf{d}_{\text{opt, pinv}} $ due to perturbations can be summarized as:
\begin{equation}
	\delta \mathbf{d}_{\text{opt, pinv}}  = -\mathbf{A}_c^+ \delta \mathbf{b} +\mathbf{A}_c^+ (\delta \mathbf{A}\odot\mathbf{c}+ \mathbf{A}\odot\delta\mathbf{c})\mathbf{d}_{\text{opt, pinv}}.
\end{equation}
\subsection{Interpretation}
1. Perturbation in $\mathbf{b}$:
\newline The sensitivity to changes in $\mathbf{b}$ is directly proportional to the pseudo-inverse of $\mathbf{A}_c$. Larger values in $\mathbf{A}_c^+$ will amplify the effect of changes in $\mathbf{b}$.

2. Perturbation in $A$:
\newline The sensitivity to changes in $\mathbf{A}_c$ is modulated by both the pseudo-inverse $\mathbf{A}_c^+$ (or the LSS) and the current solution $\mathbf{d}$. This indicates that the influence of changes in $\mathbf{A}_c$ on the solution $\mathbf{d}$ depends on the current state of the solution itself.

\section{Regularized LSS}
\label{sec: reg}
The least squares problem can be regularized to enhance the stability of its solution and potentially induce sparsity, albeit at the expense of some performance, as seen in techniques like Ridge, LASSO, or by imposing constraints on the magnitude of its components to ensure passiveness.
\subsection{Ridge Regression}
To solve the problem with regularization using ridge regression (also called Tikhonov regularization), the  objective function $ f(\mathbf{d},\lambda) $ can be expressed as a ridge-regularized version of the least squares problem
\begin{equation}
	f_{\text{ridge}}(\mathbf{d})= \| \mathbf{b} + \mathbf{A}_c\mathbf{d} \|^2 + \lambda \| \mathbf{d} \|^2,
\end{equation}
where $ \lambda > 0 $ is a regularization parameter.

As shown in Appendix \ref{app:ridge}, the ridge-regularized solution for $\mathbf{d}$, denoted as $ \mathbf{d}_{\text{opt, ridge}} $, is:
\begin{equation}
	\mathbf{d}_{\text{opt, ridge}} = - (\mathbf{A}_c^\dagger \mathbf{A}_c + \lambda\mathbf{I})^{-1} \mathbf{A}_c^\dagger \mathbf{b},
\end{equation}
where $ \lambda $ controls the amount of regularization.
This solution balances relying on the CSI (through $ \mathbf{b} + \mathbf{A}\mathbf{D}\mathbf{c} $) and accounting for possible perturbations and accordingly controlling high sensitivity by penalizing large values of $\mathbf{D}$.

The perturbation in the ridge-regularized solution $ \mathbf{d}_{\text{opt, ridge}} $ has been derived in \eqref{eq:f_sens_ridge}, the result is as follows:
\begin{align}
	&\delta \mathbf{d}_{\text{opt, ridge}} = - (\mathbf{A}_c^\dagger \mathbf{A}_c + \lambda\mathbf{I})^{-1} \left[ (\delta \mathbf{A}_c)^\dagger \mathbf{b} + \mathbf{A}_c^\dagger \delta \mathbf{b} \right] \notag\\
	&+ (\mathbf{A}_c^\dagger \mathbf{A}_c + \lambda\mathbf{I})^{-1} \left( \mathbf{A}_c^\dagger \delta \mathbf{A}_c + (\delta \mathbf{A}_c)^\dagger \mathbf{A}_c \right) \mathbf{d}_{\text{opt, ridge}},	
	\label{eq:sensitivity_ridge}
\end{align}
which demonstrates the trade-off between the regularization (through $ \lambda\mathbf{I} $) and the CSI-fitting term, and how both are affected by changes in the channel variation.
\subsection{Lasso Regularization }
The objective function for Lasso regression is given by:
\begin{equation}
	f_{\text{lasso}}(\mathbf{d})= \| \mathbf{b} + \mathbf{A}_c\mathbf{d} \|^2 + 2\lambda \| \mathbf{d} \|_1,
\end{equation}

As discussed in Appendix \ref{app:lasso}, due to lack of closed form solution for the LASSO regularization, ISTA can be used as described in Subsection \ref{app:ista}.
%
	%
	%
	%
As discussed in Subsection \ref{app:lasso_sensitivity}, in order to calculate its sensitivity to perturbation in channel responses, one should determine the sensitivity of the non-regularized solution ($\mathbf{d}_{\text{opt, LSS}}$) and accordingly find the soft-threshold function's effect on it.

Thus, the perturbed soft-threshold function $ S_\lambda(d_i + \delta d_i) $ can be approximated as follows:
\begin{align}
	&S_\lambda(d_i + \delta d_i) - S_\lambda(d_i) \approx
	&\begin{cases}
		\frac{d_i^*\delta d_i + d_i(\delta d_i)^*}{2|d_i|} & \text{if } |d_i| > \lambda \\
		0 & \text{if } |d_i| \leq \lambda \\
		\max(0, \frac{ \Re(d_i^*\delta d_i) }{|d_i|}) \frac{d_i}{|d_i|} & \text{if } |d_i| = \lambda
	\end{cases}.
	\label{eq:sensitivity_lasso}.\notag
\end{align}
This approach promotes sparsity in the solution which can be employed accordingly.
\subsection{Passiveness Constraint on the RIS}
The problem can be modified to consider the passive RIS configurations as the following optimization problem:
\begin{equation}
	\min_{x} \|\mathbf{A}_c\mathbf{d} +\mathbf{b}\|^2 \quad \text{s.t.} \quad |d_i|^2 \leq 1; \quad \forall i\in\{1,\dots,n\},
\end{equation}
where $ |d_i|^2 \leq 1 $ means that the magnitude of each element $ d_i $ of the vector $ \mathbf{d} $ is constrained to be less than or equal to 1.

Since the objective function is quadratic and convex, and the constraints are element-wise on $ \mathbf{d} $, this problem can be categorized as a constrained convex optimization problem.
There are well-known methods to address this problem such as the Lagrangian Method, PGD, and Interior-Point Method.

We employ the PGD since it works well when the problem is differentiable and when projection onto the feasible set is straightforward. It is typically used for fast, approximate solutions.
As shown in Alg. \ref{alg:PGD}, it uses gradient descent to minimize the objective and at each iteration, $ \mathbf{d} $ is projected onto the set $ \{\mathbf{d}: |d_i|^2 \leq 1\} $ to ensure $ |d_i| \leq 1 $ for each $ i $.
\begin{algorithm}[ht]
	\SetAlgoLined
	\KwData{$\mathbf{A}_c \in \mathbb{C}^{m \times n}$, $\mathbf{b} \in \mathbb{C}^{m \times 1}$, step size $\alpha$, tolerance $\epsilon$, max iterations $N_{\text{max}}$}
	\KwResult{Optimal solution $\mathbf{d}^*$}
	\textbf{Initialization:} Initialize $\mathbf{d} \in \mathbb{C}^{n \times 1}$;
	
	\For{$k = 1$ to $N_{\text{max}}$, untill convergence}{
		$\nabla f(\mathbf{d}) = 2 \mathbf{A}_c^\dagger (\mathbf{A}_c \mathbf{d} + \mathbf{b})$\;
		
		$\mathbf{d} = \mathbf{d} - \alpha \nabla f(\mathbf{d})$\;
		
		\For{$i = 1$ to $n$}{
				${d}_{ i} = {d}_{ i} / \max({|{d}_{ i}|},{1})$\;
		}
		%
		%
	}
	\Return $\mathbf{d}^*$\;
	\caption{Projected Gradient Descent Algorithm.}
	\label{alg:PGD}
\end{algorithm}

\subsection{Real-Time Calculation Induced Complexity}
It is important to note that, in addition to analyzing the sensitivity of the solutions, perturbation analysis offers a computationally efficient method for updating the RIS values, even on edge devices. This is because it only requires a few matrix-vector multiplications - specifically, 6 multiplications for LSS and Ridge, and 5 for Pinv - eliminating the need for calculating the inverse or pseudo-inverse of matrices, along with additional multiplications.
\section{Simulation and Discussion}
\label{sec: sim}
In this section, the signal level at the receiver ($S$) is used as the performance metric for simulation outputs, with results presented in terms of SNR and perturbations.

Considering the main model (\eqref{eq:main model}: $	\mathbf{y}_r=  (\mathbf{b} + \mathbf{A}\mathbf{D}\mathbf{c})\mathbf{x}_t+\mathbf{w}_r$),
the transmitted signal is normalized to one and sent through the channel $\mathbf{b}$. The elements of $\mathbf{b}$ are independently drawn from a truncated complex zero-mean normal distribution with unit variance, ensuring $\|\mathbf{b}\|_2 \leq 1$. The additive noise is modeled as $\mathbf{w}_r \sim \mathcal{CN}(0,\,\sigma^2)$, where the logarithm of $\sigma^2$ represents the SNR value at the receiver.
The components of $\mathbf{A}$ and $\mathbf{c}$ are generated in the same way to ensure the beam experiences a passive channel, with each column of $\mathbf{A}$ normalized to one.

The signal level at the receiver, $S =\|(\mathbf{b} + \mathbf{A}\mathbf{D}\mathbf{c})\mathbf{x}_t\|^2$, is reported. To evaluate the algorithm's performance in the presence of a perturbed channel or imperfect CSI, each perturbation term ($\delta \mathbf{A}$, $\delta \mathbf{b}$, and $\delta \mathbf{c}$) is randomly generated, with its norm constrained to a specified value ($\sigma_p$).
\subsection{Perfect CSI}
\begin{figure}[!hbt]
	\centering
	\includegraphics[width=0.8\textwidth]{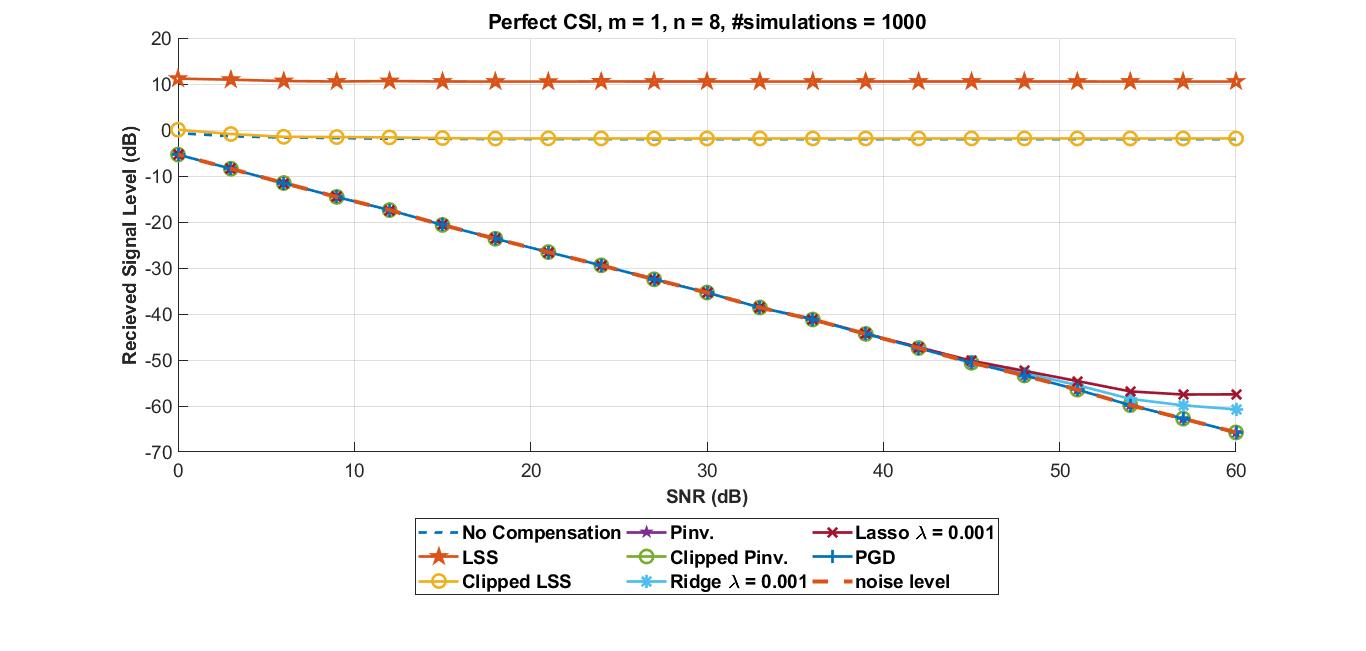}
	\caption{Rx. Signal level vs. SNR, $(m,n) = (1,8)$.}
	\label{fig:perfectCSI2}
\end{figure}

\begin{figure}[!hbt]
	\centering
	\includegraphics[width=0.8\textwidth]{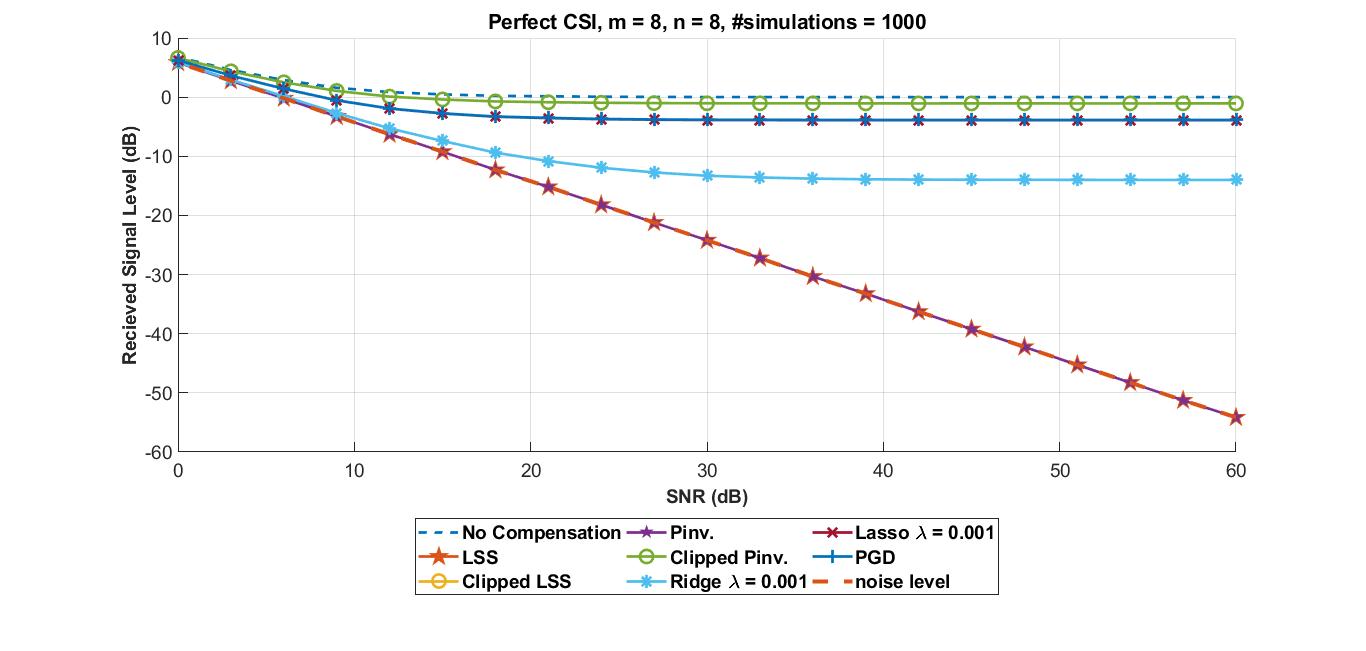}
	\caption{Rx. Signal level vs. SNR, $(m,n) = (8,8)$.}
	\label{fig:perfectCSI4}
\end{figure}

\begin{figure}[!hbt]
	\centering
	\includegraphics[width=0.8\textwidth]{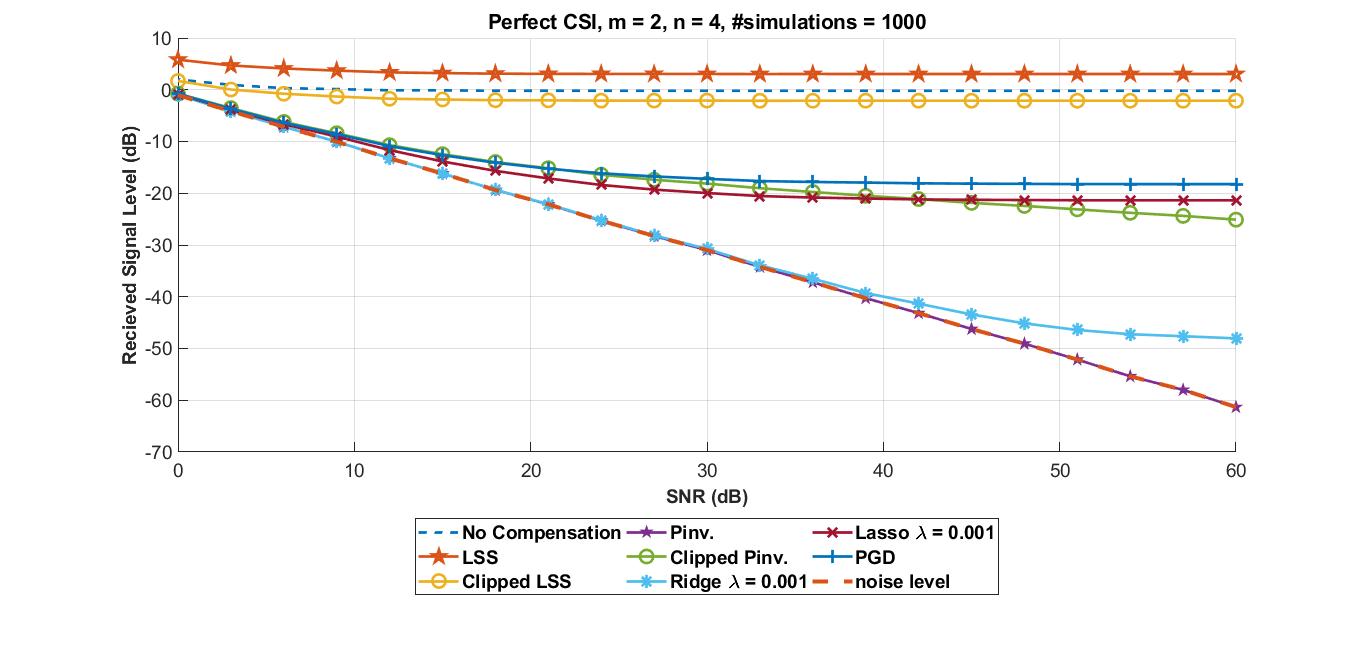}
	\caption{Rx. Signal level vs. SNR, $(m,n) = (2,4)$.}
	\label{fig:perfectCSI}
\end{figure}

First, as illustrated in Fig. \ref{fig:perfectCSI2}, \ref{fig:perfectCSI4}, and \ref{fig:perfectCSI}, we consider the ideal scenario where the perfect CSI is available. As expected, the pseudo-inverse (Pinv) shows the best performance, as it attempts to minimize the signal level at the receiver, as shown in Fig. \ref{fig:perfectCSI2}. On the other hand, the LSS approach is unsuitable for scenarios where $m < n$, as shown in Fig. \ref{fig:perfectCSI2}.

It is also worth noting that an increase in the number of receivers provides a diversity gain for the receiver, making it more difficult to achieve lower signal levels. A comparison between the pseudo-inverse and its clipped version (denoted as "Clipped Pinv.") highlights the need for an active RIS to mitigate the impact of this diversity gain at the receiver (Fig. \ref{fig:perfectCSI4} and \ref{fig:perfectCSI}).
It can be seen that the performance of the algorithms is similar at relatively low SNR values. However, as the SNR increases, the pseudo-inverse outperforms the other methods.

By adding the signal level to the SNR value, one can calculate the resulting SNR that the detector can achieve at the receiver. For example, as shown in Fig. \ref{fig:perfectCSI2}, at an SNR value of 20dB, the detector's SNR degrades to $-5$dB, indicating a 25dB reduction in the signal level at the receiver.
\subsection{Perturbed Channel Estimation (Imperfect CSI)}
\begin{figure}[!hbt]
	\centering
	\includegraphics[width=0.8\textwidth]{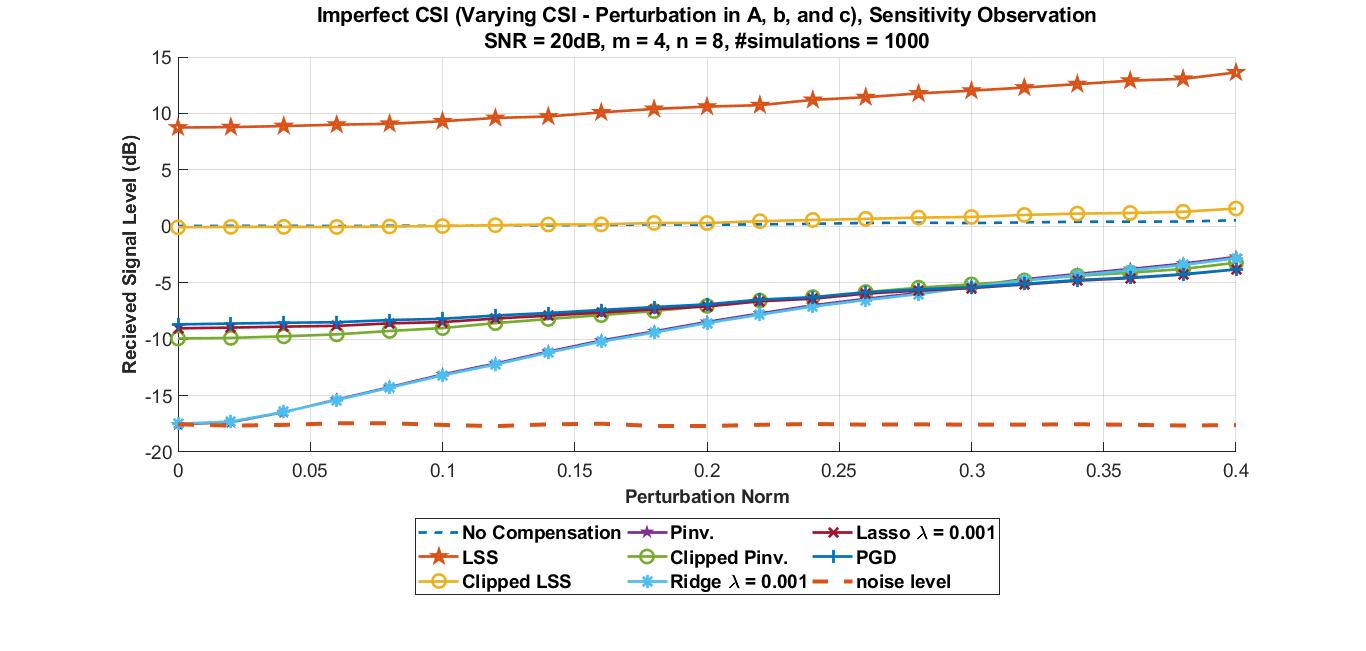}
	\caption{Rx. Signal level vs. $\sigma_p$, $(m,n) = (4,8)$.}
	\label{fig:imperfectCSI2}
\end{figure}

\begin{figure}[!hbt]
	\centering
	\includegraphics[width=0.8\textwidth]{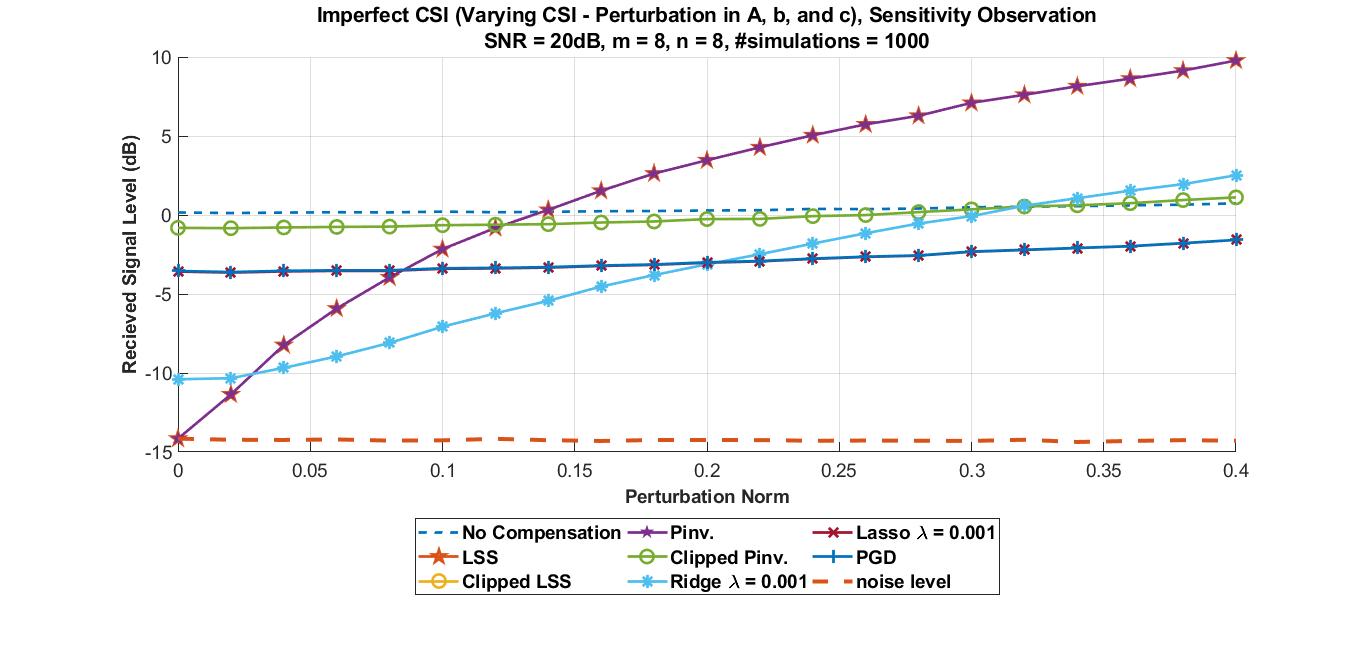}
	\caption{Rx. Signal level vs. $\sigma_p$, $(m,n) = (8,8)$.}
	\label{fig:imperfectCSI3}
\end{figure}

To study the effect of imperfect CSI on the performance of the algorithms with respect to perturbation norm, we consider various scenarios at an SNR of 20dB, as depicted in Fig. \ref{fig:imperfectCSI2} and \ref{fig:imperfectCSI3}.

It is evident that imperfect CSI degrades performance, and as previously discussed, the pseudo-inverse and, more broadly, the LSS approach become increasingly sensitive to CSI perturbations when the number of receivers equals the number of RIS elements ($m=n$). This heightened sensitivity makes the system more vulnerable to inaccuracies in channel estimation under these conditions.
\subsection{First-order Correction}
To address the performance degradation caused by imperfect CSI, we focus on the pseudo-inverse and Ridge regularization techniques. We employ the derived first-order approximation, as presented in Sections \ref{sec: LS} and \ref{sec: reg}, to correct the estimations.
\begin{figure}[!hbt]
	\begin{subfigure}[b]{0.8\textwidth}
		\includegraphics[width=\textwidth]{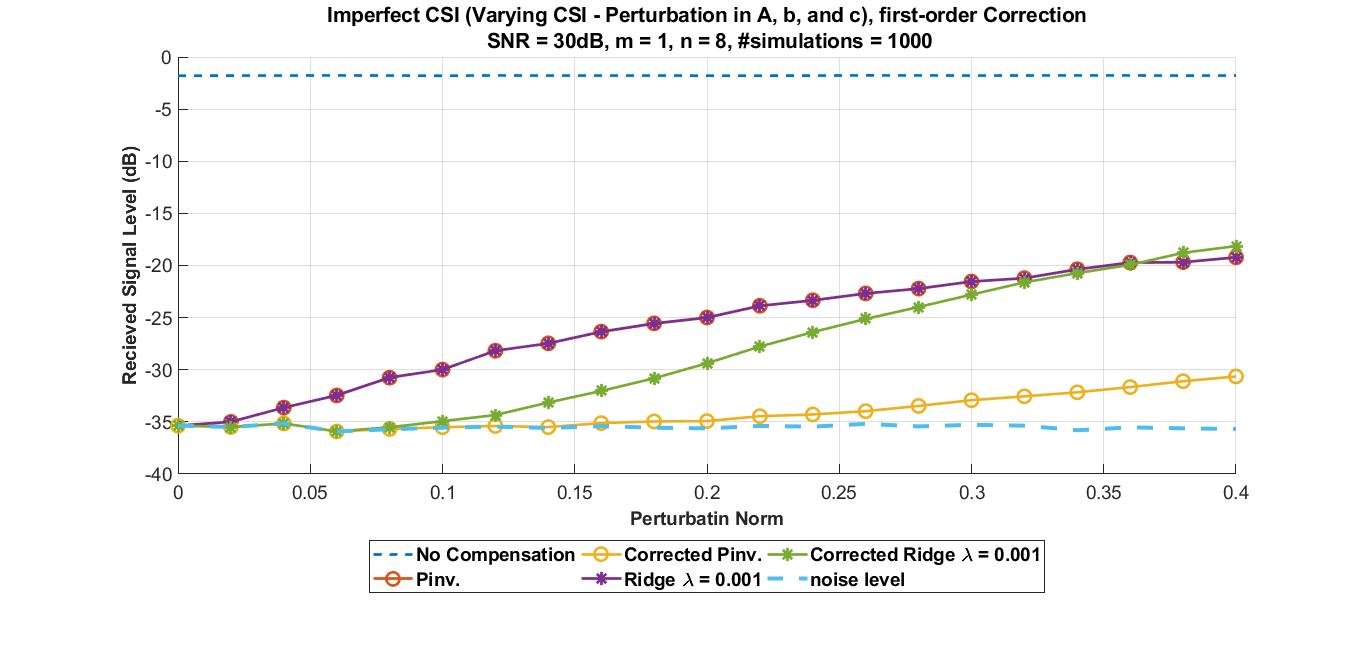}
		\subcaption{
			$(m,n) = (1,8)$.
		}
	\end{subfigure}
	\par\bigskip 
	\begin{subfigure}[b]{0.8\textwidth}
		\includegraphics[width=\textwidth]{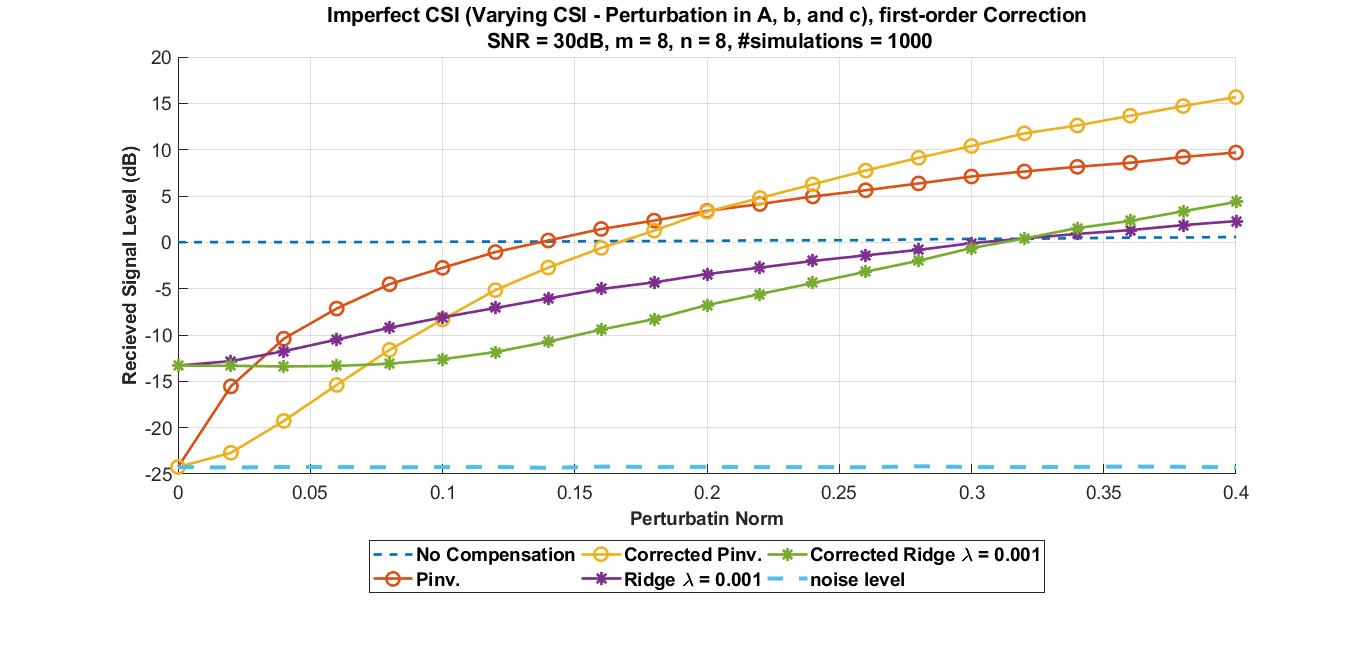}
		\subcaption{$(m,n) = (8,8)$.}
	\end{subfigure}
	\caption{
		Rx. Signal level vs. $\sigma_p$, Imperfect CSI, $SNR=30dB$.}
	\label{fig:1stOrderCorr}
\end{figure}

The results show how sensitive each approach is to perturbations, and how the first-order correction significantly expands the stability range of the algorithms, as illustrated in Fig. \ref{fig:1stOrderCorr}. Additionally, it is worth noting that using the first-order correction is much more efficient than reapplying the algorithm from scratch for each new CSI estimation, as it reduces the computational cost associated with handling the estimated perturbation.
\subsection{Sensitivity and first-order Approximation}
\begin{figure}[!hbt]
	\centering
	\includegraphics[width=0.8\textwidth]{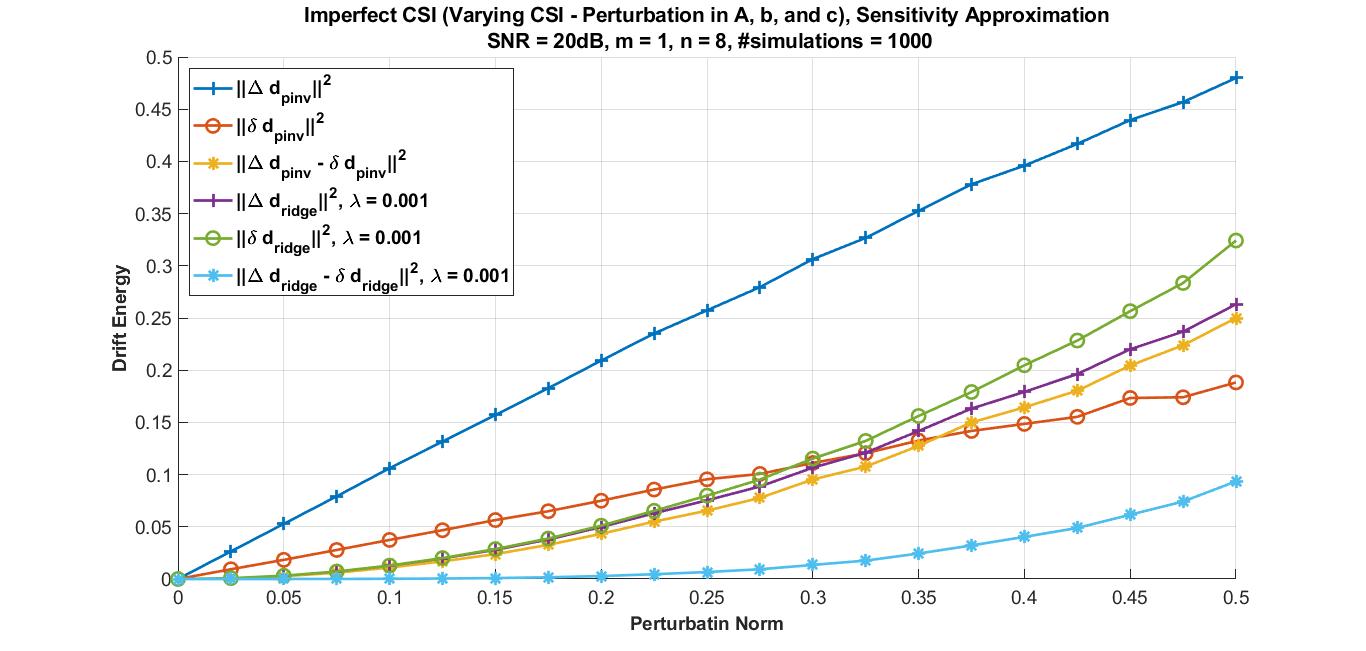}
	\caption{Drift energy vs. $\sigma_p$, $SNR=20dB$, $(m,n) = (1,8)$.}
	\label{fig:sensitivity_Abc}
\end{figure}

\begin{figure}[!hbt]
	\centering
	\includegraphics[width=0.8\textwidth]{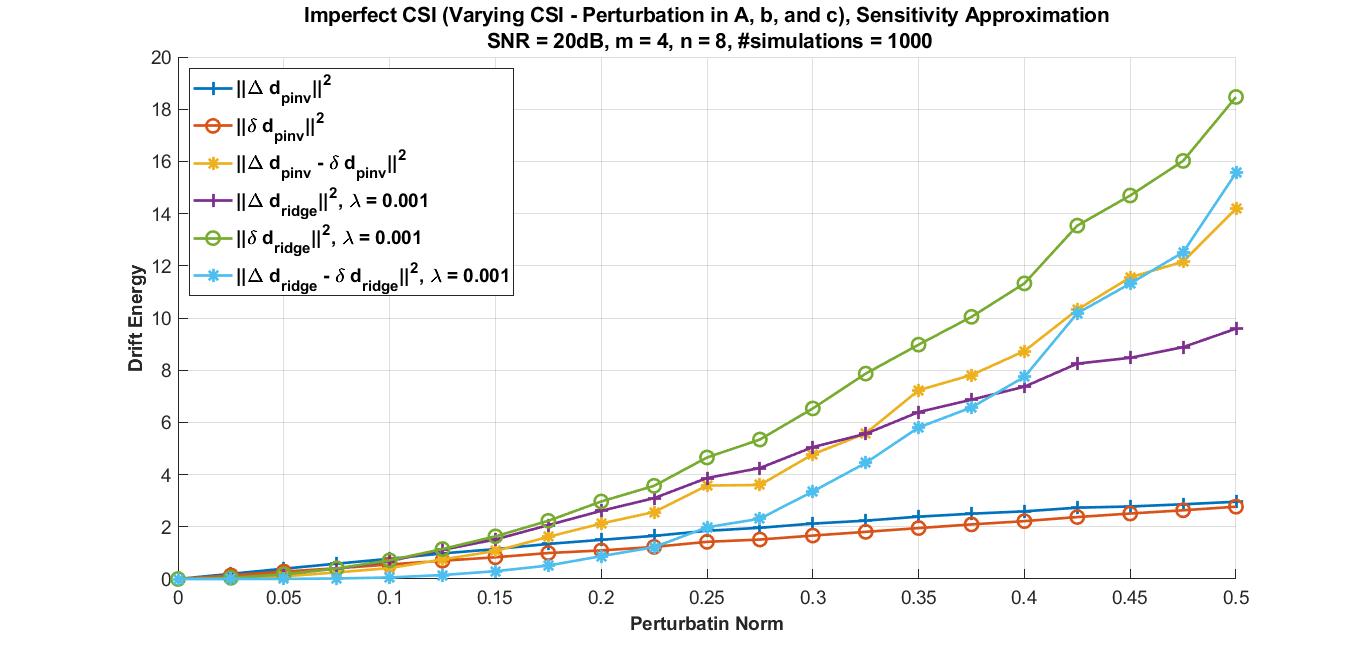}
	\caption{Drift energy vs. $\sigma_p$, $SNR=20dB$, $(m,n) = (4,8)$.}
	\label{fig:sensitivity_Abc2}
\end{figure}

\begin{figure}[!hbt]
	\centering
	\includegraphics[width=0.8\textwidth]{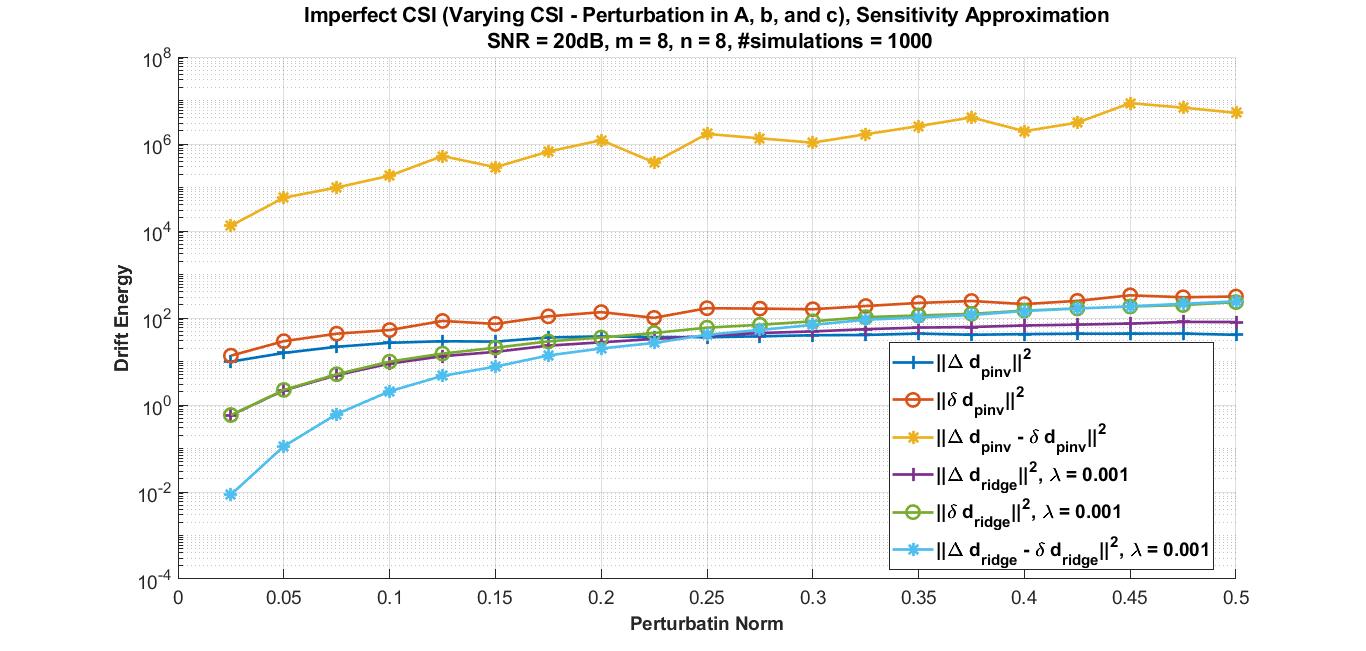}
	\caption{Drift energy vs. $\sigma_p$, $SNR=20dB$, $(m,n) = (8,8)$.}
	\label{fig:sensitivity_Abc3Log}
\end{figure}

As shown in Fig. \ref{fig:sensitivity_Abc}, \ref{fig:sensitivity_Abc2}, and \ref{fig:sensitivity_Abc3Log}, to quantify the accuracy of the first-order approximation, we compare the energy of the approximated drift ($\|\delta \mathbf{d}_\text{opt}\|^2$) with the true value ($\|\Delta\mathbf{d}_\text{opt}\|^2$) and examine their difference ($\|\Delta \mathbf{d}_\text{opt} - \delta \mathbf{d}_\text{opt}\|^2$).

The results indicate that Ridge regularization exhibits greater stability around $\mathbf{d}_\text{opt}$, with its first-order approximation showing less deviation from the true value. As shown in Fig. \ref{fig:sensitivity_Abc3Log}, it is also important to highlight that, despite its superior performance in estimating $\mathbf{d}_\text{opt}$, the pseudo-inverse method is highly sensitive to changes in CSI, making it more vulnerable to inaccuracies.
\subsection{Discussion on the Results}
The simulation results reveal that perfect CSI allows the pseudo-inverse method to effectively minimize receiver signal levels. However, with the introduction of perturbations, both the pseudo-inverse and LSS methods exhibit sensitivity, especially when the number of receivers matches the number of RIS elements. The use of first-order corrections enhances stability, with Ridge regularization outperforming the pseudo-inverse in handling channel fluctuations. These findings emphasize the importance of effective CSI estimation and correction strategies for optimizing signal reception in dynamic environments.
\section{Conclusions and Future Work}
\label{sec: con}
This paper introduces an effective approach for addressing CSI perturbations in RIS-aided communication systems by employing a first-order approximation method. Through comprehensive simulations and analyses, we demonstrate that the proposed technique significantly enhances the security and performance of communication systems, particularly in scenarios involving untrusted users and potential eavesdroppers. Our approach not only improves the resilience of RIS against CSI imperfections but also maintains computational efficiency, making it well-suited for real-time implementations in next-generation networks like 6G. By reducing the computational burden associated with continuous inverse calculations, our method is particularly advantageous for practical applications. Future research should aim to extend this framework to multi-user environments and investigate its applicability in emerging technologies, such as the IoT and smart city infrastructures.
\appendix
\section{The LSS Derivation}
\label{app:LSS}
The objective function can be expanded as:
\begin{align}
	f_{_{LSS}}(\mathbf{d}) &= \left(  \mathbf{b} + \mathbf{A}_c\mathbf{d} \right)^\dagger \left( \mathbf{b} + \mathbf{A}_c\mathbf{d} \right)\notag\\
	&=\mathbf{b}^\dagger \mathbf{b} + \mathbf{b}^\dagger \mathbf{A}_c \mathbf{d} + \mathbf{d}^\dagger \mathbf{A}_c^\dagger \mathbf{b} + \mathbf{d}^\dagger \mathbf{A}_c^\dagger \mathbf{A}_c \mathbf{d}.
\end{align}
To find the value of $\mathbf{d}$ that minimizes $f_{_{LSS}}(\mathbf{d})$, we need to take the derivative of the expression with respect to $\mathbf{d}$ and set it to zero.
The gradient of $f_{_{LSS}}(\mathbf{d})$ with respect to $\mathbf{d}$ is:
\begin{equation}
	\nabla_{\mathbf{d}} f_{_{LSS}}(\mathbf{d}) = 2 \mathbf{A}_c^\dagger \mathbf{A}_c \mathbf{d} + 2 \mathbf{A}_c^\dagger \mathbf{b}.
\end{equation}

Now, if $\mathbf{A}_c^\dagger \mathbf{A}_c$ is invertible,
the value of $\mathbf{d}$ that minimizes $f_{_{LSS}}(\mathbf{d})$ is:
\begin{equation}
	\boxed{\mathbf{d} = - (\mathbf{A}_c^\dagger \mathbf{A}_c)^{-1} \mathbf{A}_c^\dagger \mathbf{b}}.
\end{equation}
\subsection{Sensitivity to Perturbations in $ \mathbf{A}_c $ and $\mathbf{b}$}
\label{app:LSS_sensitivity}
In order to explore how $ \mathbf{d}_{\text{opt}} $ is affected by small changes in $ \mathbf{A}_c $ and $\mathbf{b}$, let $ \mathbf{A}_c\to \mathbf{A}_c + \delta \mathbf{A}_c $ and $ \mathbf{b}\to\mathbf{b} + \delta \mathbf{b} $. The perturbed solution becomes:
\begin{equation}
	\mathbf{d}_{\text{opt}} + \delta \mathbf{d}_{\text{opt}} = - ((\mathbf{A}_c + \delta \mathbf{A}_c)^\dagger (\mathbf{A}_c + \delta \mathbf{A}_c))^{-1} (\mathbf{A}_c + \delta \mathbf{A}_c)^\dagger (\mathbf{b} + \delta \mathbf{b}).\notag
\end{equation}
Expanding this to first order (ignoring other terms), we get:
\begin{equation}
	\mathbf{d}_{\text{opt}} + \delta \mathbf{d}_{\text{opt}} \approx - \left( (\mathbf{A}_c^\dagger \mathbf{A}_c)^{-1} + \delta ((\mathbf{A}_c^\dagger \mathbf{A}_c)^{-1}) \right) \left( \mathbf{A}_c^\dagger \mathbf{b} + \delta(\mathbf{A}_c^\dagger \mathbf{b}) \right).\notag
\end{equation}
Each term can be expanded as:
\begin{enumerate}
	\item Perturbation of $ \mathbf{A}_c^\dagger \mathbf{A}_c $:
	\newline The perturbation of $ \mathbf{A}_c^\dagger \mathbf{A}_c $ is:
	\begin{align}
		\delta(\mathbf{A}_c^\dagger \mathbf{A}_c) &= (\mathbf{A}_c + \delta \mathbf{A}_c)^\dagger (\mathbf{A}_c + \delta \mathbf{A}_c) - \mathbf{A}_c^\dagger \mathbf{A}_c.\notag\\
		& \approx \mathbf{A}_c^\dagger \delta \mathbf{A}_c + (\delta \mathbf{A}_c)^\dagger \mathbf{A}_c.
	\end{align}
	Thus, the perturbation in $ (\mathbf{A}_c^\dagger \mathbf{A}_c)^{-1} $ can be approximated using the matrix derivative:
	\begin{equation}
		\delta((\mathbf{A}_c^\dagger \mathbf{A}_c)^{-1}) \approx -(\mathbf{A}_c^\dagger \mathbf{A}_c)^{-1} \delta(\mathbf{A}_c^\dagger \mathbf{A}_c) (\mathbf{A}_c^\dagger \mathbf{A}_c)^{-1}.
	\end{equation}
	\item Perturbation of $ \mathbf{A}_c^\dagger \mathbf{b} $:
	\newline The perturbation in $ \mathbf{A}_c^\dagger \mathbf{b} $ due to changes in $ \mathbf{A} $ and $ \mathbf{b} $ is:
	\begin{equation}
		\delta(\mathbf{A}_c^\dagger \mathbf{b}) = (\delta \mathbf{A}_c)^\dagger \mathbf{b} + \mathbf{A}_c^\dagger \delta \mathbf{b}.
	\end{equation}
\end{enumerate}
Thus, we have:
\begin{align}
	\delta \mathbf{d}_{\text{opt}} = &- (\mathbf{A}_c^\dagger \mathbf{A}_c)^{-1} \left[ (\delta \mathbf{A}_c)^\dagger \mathbf{b} + \mathbf{A}_c^\dagger \delta \mathbf{b} \right]\notag\\
	&+ (\mathbf{A}_c^\dagger \mathbf{A}_c)^{-1}(\mathbf{A}_c^\dagger \delta \mathbf{A}_c + (\delta \mathbf{A}_c)^\dagger \mathbf{A}_c) (\mathbf{A}_c^\dagger \mathbf{A}_c)^{-1} \mathbf{A}_c^\dagger \mathbf{b}
	\label{eq:sensitivity_reg}
\end{align}
This expression captures the first-order sensitivity of $ \mathbf{d}_{\text{opt}} $ to perturbations in both $ \mathbf{A}_c $ and $ \mathbf{b} $.
It is worth mentioning, $\delta \mathbf{A}_c $ can be written in terms of $\delta \mathbf{A}$ and $\delta\mathbf{c}$ as:
\begin{equation}
	\delta \mathbf{A}_c  \approx\delta \mathbf{A} \odot  \mathbf{c} + \mathbf{A} \odot  \delta\mathbf{c}.
\end{equation}
\subsection{Perturbation in the cost function }
\label{app:LSS_perinf}
Expanding \eqref{eq:fLSS_perturbed} to first order, we get:
\begin{equation}
	f_{_{LSS}}(\mathbf{d}) + \delta 	f_{_{LSS}}(\mathbf{d}) = \| \mathbf{b} + \mathbf{A}\mathbf{D}\mathbf{c} + \delta \mathbf{b} + (\delta \mathbf{A})\mathbf{D} \mathbf{c}+ \mathbf{A} \mathbf{D}\delta \mathbf{c} \|^2.
\end{equation}
Using $\| x + y \|^2 = \| x \|^2 + 2 \Re(x^\dagger y) + \| y \|^2$, we can write:
\begin{align}
	\delta f_{_{LSS}}(\mathbf{d})  = 2 \Re \left( (\mathbf{b} + \mathbf{A}_c\mathbf{d})^\dagger (\delta \mathbf{b} + \delta \mathbf{A}_c\mathbf{d}) \right)
	+ \|\delta \mathbf{b} + \delta \mathbf{A}_c\mathbf{d} \|^2.
	\label{eq:f_sens_LSS}
\end{align}
Thus, first-order term is the linear perturbation term and consists of three components:
\begin{itemize}
	\item The contribution from $ \delta \mathbf{b} $:
	$2 \Re \left( (\mathbf{b} + \mathbf{A}_c\mathbf{d})^\dagger \delta \mathbf{b} \right)$.
	\item The contribution from $ \delta \mathbf{A} $:
	$2 \Re \left( (\mathbf{b} + \mathbf{A}\mathbf{D}\mathbf{c})^\dagger (\delta \mathbf{A} \odot\mathbf{c})\mathbf{d}\right)$.
	\item The contribution from $ \delta \mathbf{c} $:
	$2 \Re \left( (\mathbf{b} + \mathbf{A}\mathbf{D}\mathbf{c})^\dagger (\mathbf{A}\odot\delta\mathbf{c})\mathbf{d} \right)$.
\end{itemize}
\section{Pseudo-inverse Derivation}
\label{app:pinv}
The system of equations $\mathbf{A}_c\mathbf{d}+\mathbf{b}=0$ can be solved as:
\begin{equation}
	\boxed{\mathbf{d}_{\text{opt, pinv}} = -\mathbf{A}_c^+ \mathbf{b}}.
\end{equation}

- If $ \mathbf{A}_c $ has full rank, $\mathbf{A}_c^+ = (\mathbf{A}_c^\dagger \mathbf{A}_c)^{-1} \mathbf{A}_c^\dagger  $, which is the usual least-squares solution.

- If $ \mathbf{A}_c $ is rank-deficient or not full rank, the pseudo-inverse $\mathbf{A}_c^+ $ provides the best approximation to solving the system, even when $ \mathbf{A}_c^\dagger \mathbf{A}_c $ is singular.

For a matrix $ \mathbf{A}_c  $, the pseudo-inverse can be computed using a singular value decomposition (SVD):

1. Decompose $ \mathbf{A}_c  $ as $ \mathbf{A}_c  = U \Sigma V^\dagger $, where we have:

- $ U $ and $ V $: unitary matrices (in the complex case),

- $ \Sigma $: a diagonal matrix containing the singular values of $ A $.

2. The pseudo-inverse of $ \mathbf{A}_c  $ is then given by $\mathbf{A}_c ^+ = V \Sigma^+ U^\dagger$,
where $ \Sigma^+ $ is the diagonal matrix formed by taking the reciprocals of the non-zero singular values in $ \Sigma $ (and leaving the zero entries unchanged).
\subsection{Pseudo-inverse Sensitivity}
\label{app:pinv_sensitivity}
Given the perturbed matrix $ \mathbf{A}_c +\delta \mathbf{A}_c $ and the perturbed vector $\mathbf{b} + \delta \mathbf{b}$, the modified equation for the least-squares problem is:
\begin{equation}
	( \mathbf{A}_c +\delta \mathbf{A}_c ) \mathbf{d}' + (\mathbf{b} + \delta \mathbf{b}) = 0
\end{equation}
Thus, the perturbed solution $\mathbf{d}'$ can be found using the pseudo-inverse approach:
\begin{equation}
	\mathbf{d}' = - ( \mathbf{A}_c +\delta \mathbf{A}_c )^+ (\mathbf{b} + \delta \mathbf{b})
\end{equation}
To analyze how $\mathbf{d}$ changes, we will first need to express the pseudo-inverse of the perturbed matrix $ \mathbf{A}_c +\delta \mathbf{A}_c $. The first-order approximation of the pseudo-inverse can be challenging to derive directly. However, we can express it in terms of small perturbations.

Using the identity for the derivative of the pseudo-inverse, we have:
\begin{equation}
	( \mathbf{A}_c +\delta \mathbf{A}_c )^+ \approx\mathbf{A}_c^+ -\mathbf{A}_c^+ \delta \mathbf{A}_c \mathbf{A}_c^+
\end{equation}

This approximation holds when the perturbations are small and when $\mathbf{A}_c$ is full rank (or at least of maximal rank).

In order to derive the approximation for the pseudoinverse of a perturbed matrix $(\mathbf{A}_c + \delta \mathbf{A}_c)^+$ using a first-order perturbation expansion, we assume that the perturbation $\delta \mathbf{A}_c$ is small compared to $\mathbf{A}_c$, and we use the properties of the Moore-Penrose pseudoinverse.

The pseudoinverse $\mathbf{A}_c^+$ of a matrix $\mathbf{A}_c$ satisfies the following identities (Moore-Penrose conditions):
\begin{equation}
	\mathbf{A}_c \mathbf{A}_c^+ \mathbf{A}_c = \mathbf{A}_c, \quad \mathbf{A}_c^+ \mathbf{A}_c \mathbf{A}_c^+ = \mathbf{A}_c^+
\end{equation}

Now consider the perturbed matrix $ \mathbf{A}_c + \delta \mathbf{A}_c $. The pseudoinverse of this perturbed matrix should approximately satisfy the same conditions:
\begin{equation}
	(\mathbf{A}_c + \delta \mathbf{A}_c)(\mathbf{A}_c + \delta \mathbf{A}_c)^+ (\mathbf{A}_c + \delta \mathbf{A}_c) \approx \mathbf{A}_c + \delta \mathbf{A}_c
\end{equation}

Assume that the pseudoinverse of the perturbed matrix $(\mathbf{A}_c + \delta \mathbf{A}_c)^+$ can be written as a first-order approximation in $\delta \mathbf{A}_c$:
\begin{equation}
	(\mathbf{A}_c + \delta \mathbf{A}_c)^+ \approx \mathbf{A}_c^+ + \delta \mathbf{A}_c^+,
\end{equation}
where $\delta \mathbf{A}_c^+$ is a small correction term that we need to find.

Substitute this approximation into the pseudoinverse condition for the perturbed matrix:
\begin{equation}
	(\mathbf{A}_c + \delta \mathbf{A}_c) \left( \mathbf{A}_c^+ + \delta \mathbf{A}_c^+ \right) (\mathbf{A}_c + \delta \mathbf{A}_c) \approx \mathbf{A}_c + \delta \mathbf{A}_c
\end{equation}

Expand this product, keeping terms up to first order in $\delta \mathbf{A}_c$ and $\delta \mathbf{A}_c^+$ (neglect higher-order terms):
\begin{align}
	\mathbf{A}_c + \delta \mathbf{A}_c \approx \mathbf{A}_c \mathbf{A}_c^+ \mathbf{A}_c + \mathbf{A}_c \delta\mathbf{A}_c^+ \mathbf{A}_c+ \mathbf{A}_c \mathbf{A}_c^+ \delta\mathbf{A}_c
	+ \delta  \mathbf{A}_c\mathbf{A}_c^+\mathbf{A}_c.\notag
\end{align}

From the pseudoinverse property of $\mathbf{A}_c$, we know:
\begin{equation}
	\mathbf{A}_c \mathbf{A}_c^+ \mathbf{A}_c = \mathbf{A}_c.
\end{equation}
Substitute this into the expression:
\begin{equation}
	\mathbf{A}_c  + \mathbf{A}_c \delta\mathbf{A}_c^+ \mathbf{A}_c+ \mathbf{A}_c \mathbf{A}_c^+ \delta\mathbf{A}_c
	+ \delta  \mathbf{A}_c\mathbf{A}_c^+\mathbf{A}_c \approx \mathbf{A}_c + \delta \mathbf{A}_c.
\end{equation}
Subtract $\mathbf{A}_c$ from both sides:
\begin{equation}
	\mathbf{A}_c \delta\mathbf{A}_c^+ \mathbf{A}_c+ \mathbf{A}_c \mathbf{A}_c^+ \delta\mathbf{A}_c
	+ \delta  \mathbf{A}_c\mathbf{A}_c^+\mathbf{A}_c\approx \delta \mathbf{A}_c
\end{equation}
Neglect higher-order terms involving $\delta \mathbf{A}_c \delta \mathbf{A}_c^+$. We are left with the equation:
\begin{equation}
	\mathbf{A}_c \delta\mathbf{A}_c^+ \mathbf{A}_c+ \mathbf{A}_c \mathbf{A}_c^+ \delta\mathbf{A}_c
	+ \delta  \mathbf{A}_c\mathbf{A}_c^+\mathbf{A}_c= \delta \mathbf{A}_c
\end{equation}

Now solve for $\delta \mathbf{A}_c^+$. The correction term $\delta \mathbf{A}_c^+$ that satisfies this equation is:
\begin{equation}
	\delta \mathbf{A}_c^+ = -\mathbf{A}_c^+ \delta \mathbf{A}_c \mathbf{A}_c^+
\end{equation}

Thus, the first-order approximation for the pseudoinverse of the perturbed matrix $\mathbf{A}_c + \delta \mathbf{A}_c$ is:
\begin{equation}
	(\mathbf{A}_c + \delta \mathbf{A}_c)^+ \approx \mathbf{A}_c^+ - \mathbf{A}_c^+ \delta \mathbf{A}_c \mathbf{A}_c^+
\end{equation}
This is the desired result, which gives the pseudoinverse of the perturbed matrix as a first-order correction to the pseudoinverse of $\mathbf{A}_c$.

Substituting the perturbed pseudo-inverse into the equation for $\mathbf{d}'$:
\begin{equation}
	\mathbf{d}' = -\left(\mathbf{A}_c^+ -\mathbf{A}_c^+ \delta \mathbf{A}_c \mathbf{A}_c^+ \right) (\mathbf{b} + \delta \mathbf{b})
\end{equation}
\begin{equation}
	= -\mathbf{A}_c^+ \mathbf{b} -\mathbf{A}_c^+ \delta \mathbf{b} +\mathbf{A}_c^+ \delta \mathbf{A}_c \mathbf{A}_c^+ \mathbf{b}
\end{equation}
Considering $\mathbf{d} = -\mathbf{A}_c^+ \mathbf{b}$, the unperturbed solution, we have:
\begin{equation}
	\mathbf{d}' = \mathbf{d} -\mathbf{A}_c^+ \delta \mathbf{b} +\mathbf{A}_c^+ \delta \mathbf{A}_c  \mathbf{d}.
	\label{eq:f_sens_pinv}
\end{equation}
- The term $-\mathbf{A}_c^+ \delta \mathbf{b}$ indicates that a perturbation in $\mathbf{b}$ directly influences the solution linearly.
\newline - The term $\mathbf{A}_c^+ \delta \mathbf{A}_c  \mathbf{d}$ shows how the perturbation in $A$ influences $\mathbf{d}$, also linearly but scaled by $\mathbf{d}$.
\section{The Ridge Regularized Derivation}
\label{app:ridge}
Expanding the Objective function, we get:
\begin{align}
	f_{\text{ridge}}(\mathbf{d}) &= \left(  \mathbf{b} + \mathbf{A}_c\mathbf{d} \right)^\dagger \left( \mathbf{b} + \mathbf{A}_c\mathbf{d} \right) + \lambda \mathbf{d}^\dagger\mathbf{d}\notag\\
	&=\mathbf{b}^\dagger \mathbf{b} + \mathbf{b}^\dagger \mathbf{A}_c \mathbf{d} + \mathbf{d}^\dagger \mathbf{A}_c^\dagger \mathbf{b} + \mathbf{d}^\dagger \mathbf{A}_c^\dagger \mathbf{A}_c \mathbf{d} + \lambda \mathbf{d}^\dagger\mathbf{d}.\notag
\end{align}
To find the value of $\mathbf{d}$ that minimizes $f_{\text{ridge}}(\mathbf{d})$, we need to take the derivative of the expression with respect to $\mathbf{d}$ and set it to zero.
The gradient of $f_{_{LSS}}(\mathbf{d})$ with respect to $\mathbf{d}$ is:
\begin{equation}
	\nabla_{\mathbf{d}} f_{\text{ridge}}(\mathbf{d}) = 2 \mathbf{A}_c^\dagger \mathbf{A}_c \mathbf{d} + 2 \mathbf{A}_c^\dagger \mathbf{b}+2\lambda \mathbf{d}.
\end{equation}
Assuming $(\mathbf{A}_c^\dagger \mathbf{A}_c+\lambda I)$ is invertible, we can solve for $\mathbf{d}$ as:
\begin{equation}
	\boxed{\mathbf{d} = - (\mathbf{A}_c^\dagger \mathbf{A}_c+\lambda I)^{-1} \mathbf{A}_c^\dagger \mathbf{b}}.
\end{equation}
\subsection{Ridge Sensitivity}
To calculate the perturbation of the ridge-regularized solution $ \mathbf{d}_{\text{opt, ridge}} $, we will analyze how small perturbations in the matrices and vectors $ \mathbf{A} $, $ \mathbf{b} $, and $ \mathbf{c} $ affect the optimal solution vector $ \mathbf{d}_{\text{opt, ridge}} $.
The perturbed problem becomes:
\begin{align}
	&\mathbf{d}_{\text{opt, ridge}} + \Delta \mathbf{d}_{\text{opt, ridge}} =\notag\\
	&- ((\mathbf{A}_c + \delta \mathbf{A}_c)^\dagger (\mathbf{A}_c + \delta \mathbf{A}_c)+ \lambda\mathbf{I})^{-1}
	(\mathbf{A}_c + \delta \mathbf{A}_c)^\dagger (\mathbf{b} + \delta \mathbf{b}).\notag
\end{align}
Here, $ \delta \mathbf{A}_c = \delta \mathbf{A} \odot  \mathbf{c}+ \mathbf{A} \odot \delta \mathbf{c} $, which is the perturbation in $ \mathbf{A}_c $ caused by small changes in $ \mathbf{A} $ and $ \mathbf{c} $.
\subsection{Step 2: Expand the Perturbed Solution}
Expand the expression for $ \mathbf{d}_{\text{opt, ridge}} + \delta \mathbf{d}_{\text{opt, ridge}} $ to first order (ignoring higher-order terms):
\begin{align}
	&\mathbf{d}_{\text{opt, ridge}} + \delta \mathbf{d}_{\text{opt, ridge}}\approx  \notag\\
	& - \left( (\mathbf{A}_c^\dagger \mathbf{A}_c + \lambda\mathbf{I})^{-1}+ \delta ((\mathbf{A}_c^\dagger \mathbf{A}_c + \lambda\mathbf{I})^{-1}) \right) \left( \mathbf{A}_c^\dagger \mathbf{b} + \delta(\mathbf{A}_c^\dagger \mathbf{b}) \right).\notag
\end{align}
This contains two key terms to expand:
\begin{enumerate}
	\item The perturbation $ \delta ((\mathbf{A}_c^\dagger \mathbf{A}_c + \lambda\mathbf{I})^{-1}) $.
	\item The perturbation $ \delta(\mathbf{A}_c^\dagger \mathbf{b}) $.
\end{enumerate}
First, perturb $ \mathbf{A}_c^\dagger \mathbf{A}_c + \lambda\mathbf{I} $. Since $ \lambda\mathbf{I} $ is constant, its perturbation is zero, so we only need to perturb $ \mathbf{A}_c^\dagger \mathbf{A}_c $.
\begin{equation}
	\delta (\mathbf{A}_c^\dagger \mathbf{A}_c) = (\mathbf{A}_c + \delta \mathbf{A}_c)^\dagger (\mathbf{A}_c + \delta \mathbf{A}_c) - \mathbf{A}_c^\dagger \mathbf{A}_c.
\end{equation}
Expanding this term, we get:
\begin{equation}
	\delta (\mathbf{A}_c^\dagger \mathbf{A}_c) = \mathbf{A}_c^\dagger \delta \mathbf{A}_c + (\delta \mathbf{A}_c)^\dagger \mathbf{A}_c.
\end{equation}

Using the matrix derivative identity for the inverse, the perturbation of $ (\mathbf{A}_c^\dagger \mathbf{A}_c + \lambda\mathbf{I})^{-1} $ is given by:
\begin{equation}
	\delta((\mathbf{A}_c^\dagger \mathbf{A}_c + \lambda\mathbf{I})^{-1}) = -(\mathbf{A}_c^\dagger \mathbf{A}_c + \lambda\mathbf{I})^{-1} \delta(\mathbf{A}_c^\dagger \mathbf{A}_c) (\mathbf{A}_c^\dagger \mathbf{A}_c + \lambda\mathbf{I})^{-1}.\notag
\end{equation}
The perturbation of $ \mathbf{A}_c^\dagger \mathbf{b} $ is:
\begin{equation}
	\delta(\mathbf{A}_c^\dagger \mathbf{b}) = (\delta \mathbf{A}_c)^\dagger \mathbf{b} + \mathbf{A}_c^\dagger \delta \mathbf{b}.
\end{equation}
By Combining the derived perturbations, we get:
\begin{align}
	&\delta \mathbf{d}_{\text{opt, ridge}} =- (\mathbf{A}_c^\dagger \mathbf{A}_c + \lambda\mathbf{I})^{-1} \delta(\mathbf{A}_c^\dagger \mathbf{b}) \notag\\
	&+ (\mathbf{A}_c^\dagger \mathbf{A}_c + \lambda\mathbf{I})^{-1} \delta(\mathbf{A}_c^\dagger \mathbf{A}_c) (\mathbf{A}_c^\dagger \mathbf{A}_c + \lambda\mathbf{I})^{-1} \mathbf{A}_c^\dagger \mathbf{b}.
\end{align}
Substituting the expressions for $ \delta (\mathbf{A}_c^\dagger \mathbf{A}_c) $ and $ \delta (\mathbf{A}_c^\dagger \mathbf{b}) $, we have:
\begin{align}
	&\delta \mathbf{d}_{\text{opt, ridge}} = - (\mathbf{A}_c^\dagger \mathbf{A}_c + \lambda\mathbf{I})^{-1} [ (\delta \mathbf{A}_c)^\dagger \mathbf{b} + \mathbf{A}_c^\dagger \delta \mathbf{b} \notag\\
	&- \left( \mathbf{A}_c^\dagger \delta \mathbf{A}_c + (\delta \mathbf{A}_c)^\dagger \mathbf{A}_c \right) (\mathbf{A}_c^\dagger \mathbf{A}_c + \lambda\mathbf{I})^{-1} \mathbf{A}_c^\dagger \mathbf{b}].
	\label{eq:f_sens_ridge}
\end{align}
\section{The LASSO Regularized Solution}
\label{app:lasso}
The gradient of the objective function with respect to $\mathbf{d}$ for the non-regularized part is:
\begin{equation}
	[\nabla f_{\text{lasso}}(\mathbf{d})]_{\text{non-regularized}}= \nabla_{\mathbf{d}} f_{_{LSS}}(\mathbf{d}) = 2 \mathbf{A}_c^\dagger \mathbf{A}_c \mathbf{d} + 2 \mathbf{A}_c^\dagger \mathbf{b}
\end{equation}
For the Lasso regularization term involving complex entries, the subgradient is given by:
\begin{equation}
	\nabla \left( \lambda \sum_{i=1}^n |d_{i}| \right) = \lambda \, \frac{d_{i}}{|d_{i}|} \quad \text{if } d_{i}\neq 0,
\end{equation}
and is undefined for $ d_{i}= 0 $.
Setting the gradient to zero, we have $	2 \mathbf{A}_c^\dagger \mathbf{A}_c \mathbf{d}  +2\lambda \, \frac{\mathbf{d}}{|\mathbf{d}|} = 0$, that leads to:
\begin{equation}
	\boxed{\mathbf{A}_c^\dagger \mathbf{A}_c \mathbf{d} =-\lambda \, \frac{\mathbf{d}}{|\mathbf{d}|}}.
\end{equation}

Lasso regression does not have a closed-form solution like ridge regression. Instead, it requires specialized optimization algorithms due to the $\ell_1$ norm's non-differentiability like Coordinate Descent Method, Subgradient Descent, ADMM \cite{wang2019global}, LARS \cite{10.1214/009053604000000067}, and ISTA. Here, we mention the ISTA.
\subsection{ISTA}
\label{app:ista}
The ISTA uses soft-thresholding to handle the $\ell_1$-norm and iteratively minimizes the Lasso objective:
\begin{equation}
	f_{\text{lasso}}(\mathbf{d}) = \frac{1}{2} \| \mathbf{b} + \mathbf{A}_c \mathbf{d} \|^2 + \lambda \| \mathbf{d} \|_1
\end{equation}
This is equivalent to minimizing:
\begin{equation}
	f(\mathbf{d}) = g(\mathbf{d}) + R(\mathbf{d}, \lambda),
\end{equation}
where $ g(\mathbf{d}) = \frac{1}{2} \| \mathbf{b} + \mathbf{A}_c \mathbf{d} \|^2 $ is smooth, differentiable and the regularization term $ R(\mathbf{d}, \lambda) = \lambda \| \mathbf{d} \|_1 $ is non-smooth and convex.

ISTA updates the solution iteratively by performing the following steps (1- Gradient Computation (GC), 2- Gradient Descent and Shrinkage (GDS)):

\begin{itemize}
	\item Initialization: initial guess $\mathbf{d}^{(0)}$ (typically $\mathbf{d}^{(0)} = \mathbf{0}$).
	\item Iterative Update (At each iteration $k$):
	
	- GC: $\nabla g(\mathbf{d}^{(k)}) = \mathbf{A}_c^\top (\mathbf{A}_c \mathbf{d}^{(k)} + \mathbf{b})$
	
	- GDS: $\mathbf{d}^{(k+1)} = S_{\lambda \alpha} \left( \mathbf{d}^{(k)} - \alpha \nabla g(\mathbf{d}^{(k)}) \right)$.
	\item Stopping Criterion: $\| \mathbf{d}^{(k+1)} - \mathbf{d}^{(k)} \| < \epsilon$.
\end{itemize}
where step size $t > 0$ is chosen based on Lipschitz constant $L$ of $\nabla g(\mathbf{d})$ and the soft-thresholding function is defined as:
\begin{equation}
	S_\lambda(d_i) \triangleq
	\begin{cases}
		\left(1 - \frac{\lambda}{|d_i|}\right) d_i & \text{if } |d_i| > \lambda \\
		0 & \text{if } |d_i| \leq \lambda
	\end{cases}.
\end{equation}
\subsection{LASSO Sensitivity Analysis}
\label{app:lasso_sensitivity}
In the case of orthonormal columns of $\mathbf{A}_c$, the Lasso solution can be written as follows, Table 3.4 in \cite{hastie2009elements}:
\begin{equation}
	\mathbf{d}_{\text{opt, lasso}} = S_\lambda\left( - (\mathbf{A}_c^\dagger \mathbf{A}_c)^{-1} \mathbf{A}_c^\dagger \mathbf{b}\right),
\end{equation}
where soft thresholding is applied component-wise.

In order to calculate its sensitivity to perturbation in channel responses, one should determine the sensitivity of the non-regularized solution ($\mathbf{d}_{\text{opt, LSS}} = - (\mathbf{A}_c^\dagger \mathbf{A}_c)^{-1} \mathbf{A}_c^\dagger \mathbf{b}$) and accordingly find its effect on the soft-threshold function.
Considering
\begin{equation}
	\sqrt{(d_i + \delta d_i)^*(d_i + \delta d_i)} \approx |d_i|\sqrt{1+\frac{\delta d_i .d_i + d_i.\delta d_i }{|d_i|^2}},
\end{equation}
the magnitude $ |d_i + \delta d_i| $ can be approximated as:
\begin{equation}
	|d_i + \delta d_i|\approx|d_i|(1 + \frac{d_i^*\delta d_i + d_i(\delta d_i)^*}{2|d_i|^2}) ,
\end{equation}
and
\begin{align}
	\frac{d_i + \delta d_i}{|d_i + \delta d_i|}&\approx \frac{d_i + \delta d_i}{|d_i|} (1 - \frac{d_i^*\delta d_i + d_i(\delta d_i)^*}{2|d_i|^2})\notag\\
	&\approx \frac{d_i}{|d_i|}+\frac{\delta d_i }{2|d_i|} -\frac{d_i^2\delta d_i^*}{2|d_i|^3}.
\end{align}

Then, the perturbed soft-threshold function $ S_\lambda(d_i + \delta d_i) $ can be approximated as follows:
\begin{align}
	&S_\lambda(d_i + \delta d_i) - S_\lambda(d_i) \approx
	&\begin{cases}
		\frac{d_i^*\delta d_i + d_i(\delta d_i)^*}{2|d_i|} & \text{if } |d_i| > \lambda \\
		0 & \text{if } |d_i| \leq \lambda \\
		\max(0, \frac{ \Re(d_i^*\delta d_i) }{|d_i|}) \frac{d_i}{|d_i|} & \text{if } |d_i| = \lambda
	\end{cases}.\notag
\end{align}






\bibliographystyle{elsarticle-num}
\bibliography{bibfile}

%
%
\end{document}